\documentclass[twocolumn,twoside,groupedaddress,nofootinbib,nobalancelastpage,
nobibnotes]{revtex4}
\pdfoutput=1
\usepackage[letterpaper, portrait,margin=.6in]{geometry}
\usepackage[hyperfootnotes=false]{hyperref}
\usepackage{amsmath,amsfonts,amssymb,mathrsfs,graphicx,color}
\usepackage[squaren]{SIunits}
\usepackage{verbatim}
\usepackage{enumerate}
\usepackage{graphicx}
\usepackage{xcolor}
\usepackage{slashed}
\usepackage[utf8]{inputenc}
\usepackage[percent]{overpic}
\usepackage{natbib}
\usepackage{multirow}
\usepackage{float}
\usepackage{booktabs}
\usepackage[titletoc,title]{appendix}

\newcommand{\lorentz}{\Gamma}

\newcommand{\ie}{{\it i.e.}}
\newcommand{\eg}{{\it e.g.}}

\newcommand{\cf}{{\it cf.}}

\newcommand{\eq}{Eq.}

\newcommand{\fig}{Fig.}

\newcommand{\Figs}{Figures}
\newcommand{\Ref}{Ref.}
\newcommand{\Refs}{Refs.}
\newcommand{\Sec}{Section}

\newcommand{\App}{Appendix}

\newcommand{\Tab}{Tab.}

\newcommand{\equals}{\,=\,}

\newcommand{\equ}[1]{\eq~(\ref{equ:#1})}
\newcommand{\figu}[1]{\fig~\ref{fig:#1}}
\newcommand{\tabl}[1]{\Tab~\ref{tab:#1}}
\newcommand{\sect}[1]{\Sec~\ref{sec:#1}}

\begin{document}

\title{Neutrinos and Ultra-High-Energy Cosmic-Ray Nuclei from Blazars}

\author{Xavier Rodrigues}
\affiliation{DESY, Platanenallee 6, 15738 Zeuthen, Germany}

\author{Anatoli Fedynitch}
\affiliation{DESY, Platanenallee 6, 15738 Zeuthen, Germany}

\author{Shan Gao}
\affiliation{DESY, Platanenallee 6, 15738 Zeuthen, Germany}

\author{Denise Boncioli}
\affiliation{DESY, Platanenallee 6, 15738 Zeuthen, Germany}

\author{Walter Winter}
\affiliation{DESY, Platanenallee 6, 15738 Zeuthen, Germany}

\date{\today}

\begin{abstract}

We discuss the production of ultra-high-energy cosmic ray (UHECR) nuclei and neutrinos from blazars. We compute the nuclear cascade in the jet for both BL Lac objects and flat-spectrum radio quasars (FSRQs), and in the ambient radiation zones for FSRQs as well. By modeling representative spectral energy distributions along the blazar sequence, two distinct regimes are identified, which we call ``nuclear survival'' -- typically found in low-luminosity BL Lacs, and ``nuclear cascade'' -- typically found in high-luminosity FSRQs. We quantify how the neutrino and cosmic-ray (CR) emission efficiencies evolve over the blazar sequence, and demonstrate that neutrinos and CRs come from very different object classes. For example, high-frequency peaked BL Lacs (HBLs) tend to produce CRs, and HL-FSRQs are the more efficient neutrino emitters. This conclusion does not depend on the CR escape mechanism, for which we discuss two alternatives (diffusive and advective escape). Finally, the neutrino spectrum from blazars is shown to significantly depend on the injection composition into the jet, especially in the nuclear cascade case: Injection compositions heavier than protons lead to reduced neutrino production at the peak, which moves at the same time to lower energies. Thus, these sources will exhibit better compatibility with the observed IceCube and UHECR data.

\end{abstract}

\maketitle

\section{Introduction}
\label{sec:introduction}

In spite of the experimental efforts to measure cosmic rays (CRs) at the highest energies, \ie~above $10^{18}$ eV, their origin is not yet clear. These ultra-high-energy cosmic rays (UHECRs) are accelerated in extragalactic sources \cite{Aab:2017tyv} which are not yet resolved. The Pierre Auger Observatory has measured observables which are sensitive to chemical composition, favoring the interpretation of a mixed CR composition~\cite{Aab:2016zth} and motivating the investigation of the interactions of nuclei heavier than hydrogen in the sources and their transport through extragalactic space. An independent test can be obtained from the observations of secondary messengers produced during CR propagation: neutrinos and $\gamma$-rays; see \eg\ \Refs~\cite{Heinze:2015hhp,Supanitsky:2016gke}.
Active galactic nuclei (AGNs) are considered as sites where the acceleration of protons and nuclei might take place. In this study we investigate the possibility that blazars, \ie~AGNs whose jet points in the direction of the observer, are the sources of the UHECRs and neutrinos, which are expected to be produced through interactions of CRs with photons in the source and its surroundings.

High-energy astrophysical neutrinos, recently detected by IceCube~\cite{Aartsen:2013jdh,Aartsen:2013bka}, are the smoking-gun signature of hadronic interactions in astrophysical environments. However, the origin of these neutrinos remains unknown. Blazars, as the dominating extragalactic gamma-ray sources \cite{Ajello:2015mfa}, have long been proposed as promising neutrino emitters if the observed gamma-rays are generated in hadronic interactions~\cite{Stecker:1991vm}. Indeed, if all blazars belonged to this type, their energy budget alone would be sufficient to power the astrophysical neutrino flux observed by IceCube.  However, \Ref~\cite{Aartsen:2016lir} has limited the fraction of this contribution to the neutrino flux from blazars to 7-27\%, after a study on the spatial correlation between the directions of the IceCube neutrinos and the positions of the sampled blazars from the Fermi 2LAC catalog \cite{Fermi-LAT:2011xmf}. 
Correlation studies between IceCube neutrinos and GeV and X-ray flares from point sources, including blazars, have been performed as well, but no statistically significant result has been found~\cite{Peng:2016tfi}, except from some hints~\cite{Padovani:2016wwn,Kadler:2016ygj,Resconi:2016ggj}. A few significant GeV blazar flares, such as 3C279 and PKS B1424-418, have been found inconsistent to be of purely protonic origin from theoretical modelings~\cite{Petropoulou:2016xat,Gao:2016uld}. 
Therefore, the important remaining questions are: What fraction of the IceCube neutrino flux can be attributed to blazars, given the possibility that the blazar jet may contain a mixture of leptons, protons and nuclei; what the characteristic neutrino energies as a function of those ingredients are; and whether blazars can be the sources of the UHECRs in spite of possible neutrino constraints. This article cannot address all of these questions, but will rather focus on the neutrino and CR production efficiencies of individual sources.

The neutrino spectrum and production efficiencies are highly dependent on the spectral energy distribution (SED) of the seed photons~\cite{Murase:2014foa}. Therefore, a meaningful prediction of the diffuse neutrino spectrum from blazars must incorporate the differences in the SEDs among the blazar family. Based on the work of \cite{Fossati:1998zn}, the description of SEDs from observed blazars can be reasonably approximated by a sequence as a function of the gamma-ray luminosity $L_{\gamma}$: The brighter the source is in gamma rays, the ``redder'' the SED becomes (\ie~with the spectral peaks at lower energies).\footnote{In~\cite{Ghisellini:2017ico} this feature becomes more pronounced due to a much larger sample of Fermi-3LAC \cite{Ackermann:2015yfk} blazars.} Based on this prescription, we study the neutrino and CR emissivity for a number of representative cases out of the sequence: from low-luminosity, high-frequency peaked BL Lacs (HBL) to high-luminosity, low-frequency peaked BL Lacs (LBL) and FSRQs. As a function of both the luminosity and jet composition, we reveal which type of blazar is the best candidate for neutrino and UHECR sources. 

Our blazar model is an extended version of \Ref~\cite{Murase:2014foa} (see also \Ref~\cite{Anchordoqui:2007tn} for an earlier approach), where the ejected neutrino spectra have been calculated for a sample of blazars of different luminosities considering a pure-proton injection composition. In that model, blazar emission cannot simultaneously explain the observed diffuse PeV neutrino flux and the UHECR flux, since a different CR loading would be required to explain either one; furthermore, the diffuse neutrino spectra obtained were found to be too high compared to IceCube observations at multi-PeV energies. In comparison with that work, in the present model we include the injection of nuclei heavier than protons including all relevant radiation processes, such as photo-nuclear disintegration. The latter process refers to all interactions exhibited by nuclei below the pion production threshold, which lead to the disintegration of the nucleus into lighter isotopes and one or multiple lighter fragments (see \eg\ \Ref~\cite{Boncioli:2016lkt}). At higher energies these processes are accompanied by inelastic interactions of target photons with individual nucleons, resulting in photo-hadronic production of mesons that can subsequently decay into neutrinos. 
Given the enhanced star-forming rates in the centers of the host galaxies, it is natural to expect high abundance of these heavy elements, as is observed from the emission lines from the accretion disk (see \eg\ \Ref~\cite{Fabian:2000nu}). In this work we assume that these elements can be transported to the jet from either the accretion disk or other locations, such as in a star-in-jet scenario~\cite{Barkov:2010km}. We demonstrate that for heavier injection compositions the emitted neutrino spectrum has a lower cutoff and overall normalization, which may help solve both aforementioned conflicts. We also include a more refined discussion of possible CR escape mechanisms from the jet zone (diffusion versus advection), and show their impact on the production of secondary particles in the broad line region (BLR) and dust torus (DT) in FSRQs. 
For the sake of simplicity, we focus on pure injection compositions into the jet, ranging from protons to iron.

The paper is structured as follows: In \sect{methods}, we introduce the blazar jet model and the simulation methods. In \sect{nuclear_regimes}, we analyze the effect of the source parameters on the nuclear cascade and identify two qualitatively different nuclear disintegration regimes in the jet. Neutrino production is discussed in \sect{neutrinos}, and the CR escape mechanisms in \sect{escape}, where the emitted CR spectra are also presented. In \sect{fsrq} we introduce a model for FSRQs which includes the BLR and DT as additional radiation zones. In \sect{sequence} we apply our model to a sample of sources across the blazar sequence and calculate their neutrino and UHECR emissivity. Our conclusions are presented in \sect{conclusion}.

\section{Methods}
\label{sec:methods}

\begin{figure}[t!]
	\includegraphics[width=0.8\columnwidth]{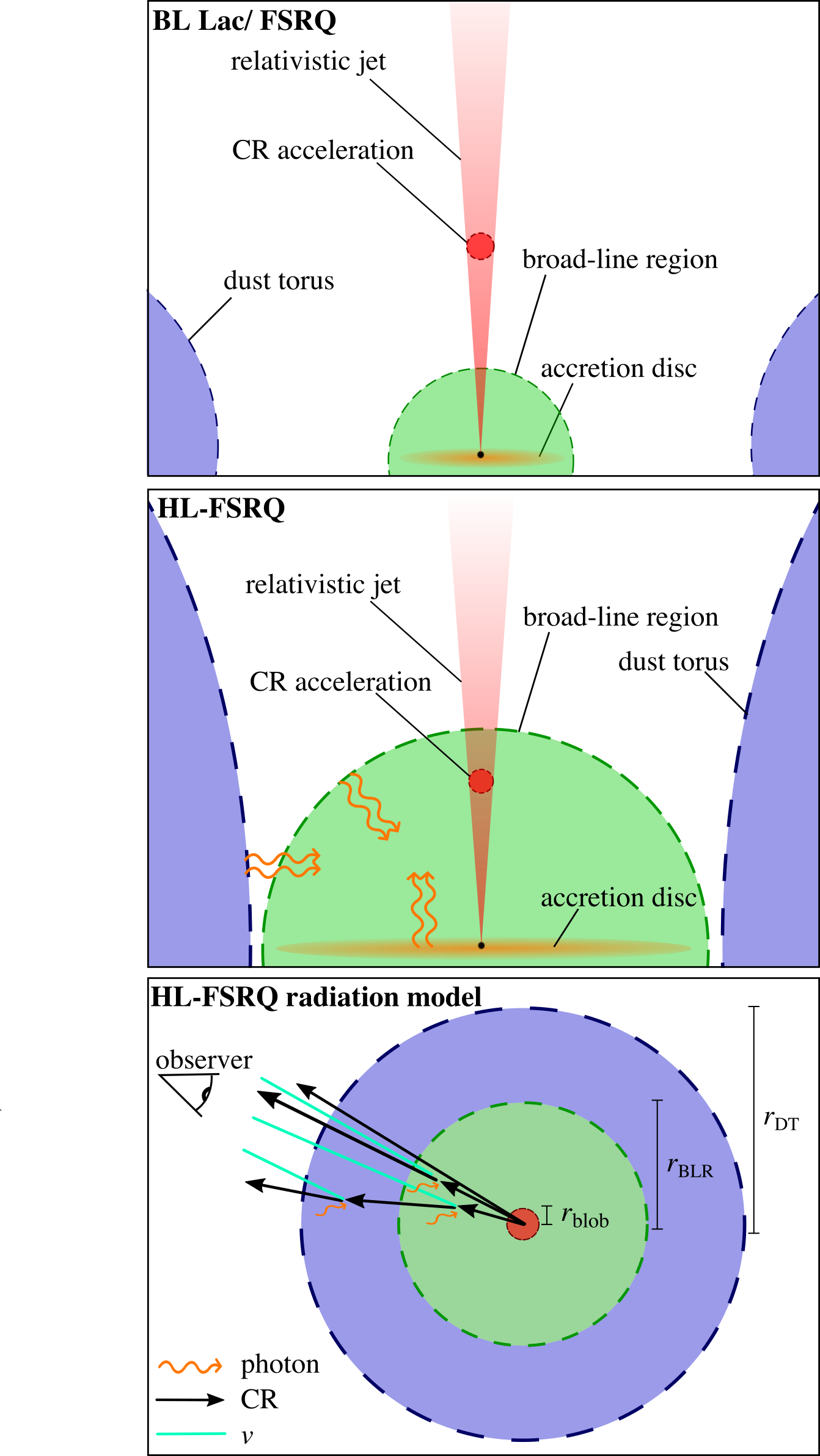}
	\caption{Schematic representation of a BL Lac scenario \textit{(top)}, in which the external radiation fields of the BLR and DT in the jet and the CR re-processing in these zones can be neglected; a high-luminosity FSRQ (HL-FSRQ) \textit{(middle)}, in which isotropized external photons contribute to the radiation fields in the jet and CRs may be re-processed in the BLR and possibly the DT; and our corresponding simplified ``spherical-cow'' radiation model of HL-FSRQs \textit{(bottom)}.}
	\label{fig:model}
\end{figure}

For the simulation of blazars with the injection of nuclei we use similar techniques as in \Refs~\cite{Boncioli:2016lkt,Biehl:2017zlw}, where details can be found in \Ref~\cite{Biehl:2017zlw}. Our jet model closely follows \Ref~\cite{Murase:2014foa} as far as geometric setup and spectral energy distributions (SEDs) are concerned. Until \sect{fsrq} we will consider only radiation processes in the jet, which includes CR interactions with the non-thermal SED, neutrino production and CR escape. In \sect{fsrq}, the model will be extended to include external radiation typically associated with FSRQs.

\subsection*{Model geometry}
\label{sec:geometry}

We assume that CRs are accelerated in a plasma blob of radius\footnote{In this work, primed quantities refer to the blob rest frame and unprimed ones to the black-hole frame or, if explicitly mentioned, to the observer's frame.} $r'_{\text{blob}} \simeq 3 \cdot 10^{16} \, \mathrm{cm}$ unless noted otherwise, where the size can be estimated from a flaring time $t'_{\text{flare}} = 10^6 \, \mathrm{s}$ such that $r'_{\text{blob}} \simeq \Gamma\, c\, t_{\text{flare}}$. The blob is spherical in its rest frame with volume $V'=(4\pi/3)r^{\prime3}_{\text{blob}}$, and travels along the jet with Doppler factor $\Gamma \simeq 10$ towards the observer.\footnote{The Lorentz factor of the jet corresponds to the Doppler factor $D$ for the observation angle $\alpha \simeq 1/\Gamma$ off the jet axis, which we assume for the sake of simplicity.} The jet is assumed to have an opening angle of $\theta \sim \lorentz^{-1} \sim 0.1$, so that the typical dissipation radius can be estimated from a simple geometry argument to be $r_{\text{diss}} \approx  \lorentz r'_{\text{blob}} = 3 \cdot 10^{17} \, \mathrm{cm}$.

The dissipation radius will not come into play until \sect{fsrq}, when we consider external radiation fields characteristic of some FSRQs, namely thermal radiation from the accretion disk and the DT, and molecular emission from the BLR. In sources where the BLR radius is larger than the jet dissipation radius, $r_{\text{BLR}}>r_{\text{diss}}$, the radiation zone of the jet will be situated within the BLR (see middle panel of \figu{model}). In this case, we will assume that the component originating from these external fields is boosted into the jet blob. In addition, the CRs that escape the jet will have to propagate through the external fields before escaping into extragalactic space from the blazar environment. In cases where the blob dissipation occurs outside the BLR ($r_{\text{diss}} > r_{\text{BLR}}$, see upper panel of \figu{model}), CRs only experience the non-thermal radiation inside the blob and those that escape the jet are effectively emitted by the source. 

We assume the radius of the BLR~\cite{Ghisellini:2008zp} and the DT~\cite{Ghisellini:2008zp,Hoenig:2007aa,Cleary:2006pe,Kishimoto:2010nu,Kishimoto:2011hz} to scale with the square-root of the accretion disk luminosity $L_{\text{disk}}$:

\begin{align}
	r_{\text{BLR}} &= 10^{17}  \left(\frac{L_{\text{disk}}}{10^{45} \mathrm{erg/s}}\right)^{1/2} ~\mathrm{cm} \,, \label{equ:disk_luminosity} \\
	r_{\text{DT}} &= 2.5 \cdot 10^{18}  \left(\frac{L_{\text{disk}}}{10^{45} \mathrm{erg/s}}\right)^{1/2} ~\mathrm{cm} \,. \label{equ:dt_luminosity}
\end{align}

Following the previous discussion, \equ{disk_luminosity} implies that CR interactions with external fields is relevant only for FSRQs with $L_{\text{disk}}>9\times10^{45}~\mathrm{erg/s}$. These sources will be denoted high-luminosity FSRQs (HL-FSRQs), as indicated in \figu{model}. For other FSRQs, as well as for BL Lacs, a single plasma blob scenario containing only non-thermal jet emission characterizes the interaction environment. In the next sections and until \sect{fsrq} we will focus on this simplified one-zone scenario. A further description of the three-zone FSRQ model will be given in \sect{fsrq}.

The bolometric luminosity of the disk, $L_\mathrm{disk}$, is roughly proportional to that of the jet, $L_\mathrm{jet}$, based on the phenomenological relationship from \Refs~\cite{Inoue:2008pk,Inoue:2010xb,Abazajian:2010pc,Lusso:2009nq}. In this work, the blazars are represented by $L_{\gamma}$, the integrated luminosity of the SED above 100 MeV. For the relationships between $L_\mathrm{disk}$, $L_\mathrm{jet}$ and $L_{\gamma}$, we follow the tabulated values in \Ref~\cite{Murase:2014foa}, where for FSRQs $L_\mathrm{disk}$ can be empirically related to $L_{\gamma}$ as $L_{\gamma} \, [\mathrm{erg \, s^{-1}}] \approx 10^{17} \,  (L_\mathrm{disk} [\mathrm{erg \, s^{-1}}])^{0.683}$.

\subsection*{Particle interactions}
\label{sec:jet_model}

\begin{figure}[t!]
 \begin{center}
 \includegraphics[width=\columnwidth]{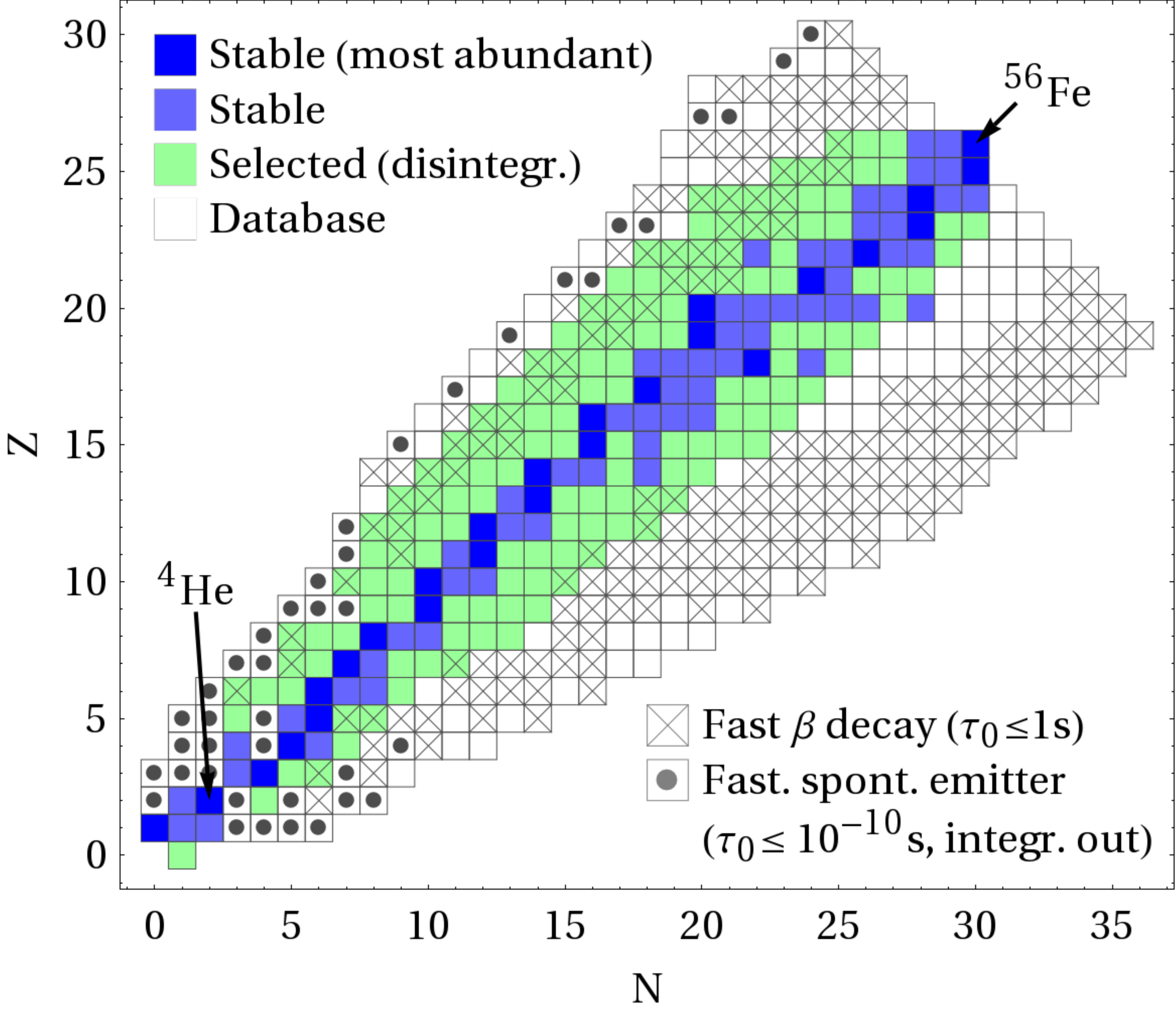}
 \end{center}
\caption{\label{fig:isotopes} Nuclear isotopes considered in this work as a function of neutron number $N$ and proton number $Z$ (``database'', white boxes). The fast spontaneous emitters marked by dots are integrated out immediately, while the colored isotopes are selected by a recursive algorithm following the leading disintegration and decay paths. The blue isotopes are stable, while the green ones are typically unstable $\beta^\pm$ emitters. Fast $\beta^\pm$ emitters (with lifetimes such that they could potentially contribute to neutrino production in the jet if interactions do not dominate) are marked with crosses, while in the dust torus basically most green-marked isotopes can be relevant.}
\end{figure}

The transport of accelerated nuclei in the source is modeled as a coupled partial differential equation (PDE) system including the colored species in \figu{isotopes}. Details may be found in \Ref~\cite{Biehl:2017zlw}, where the same method is used for solving the transport equations based on the \textsc{NeuCosmA} code~\cite{Baerwald:2011ee}. The PDE reads for the isotope $i$
\begin{equation}
\frac{\partial N'_i}{\partial t'} = \frac{\partial}{\partial E'} \left( - b'(E') N'_i(E') \right) - \frac{N'_i(E')}{t'_{\text{esc}}} + \tilde Q'_{ji}(E) \, , \label{equ:master}
\end{equation}
where $b'(E) = E' {t'}^{-1}_{\text{loss}}$ (with the energy loss rate ${t'}^{-1}_{\text{loss}}$),  and ${t'}^{-1}_{\text{esc}}$ is the escape rate. The PDE system  is to be solved for the differential particle densities $N'_i \, [ \text{GeV}^{-1} \text{cm}^{-3} ]$ in the blob frame. The coupling of the PDEs arises because of the injection term 
\begin{equation}
 \tilde Q'_{ji}(E)= Q'_i(E) + Q'_{j \rightarrow i}(E) \, , \label{equ:inject}
\end{equation}
which allows for injection from an acceleration zone $Q'_i$, as well as for injection from other species $j$ with the term  $Q'_{j \rightarrow i}$, such as from  photo-disintegration or $\beta^\pm$ decays. 
In this work, we consider pure injection compositions, which means that there is only one injection term $Q'_i$ for the injected species $i$ (such as protons, helium-4 or iron-56): 

\begin{equation}
Q'_i \propto (E_i')^{-2} \cdot \exp \left(  - \left( E_i'/E'_{i,\mathrm{max}} \right)^2 \right).
\label{equ:inj} 
\end{equation}

The maximum energy $E'_{i,\text{max}}$ is determined by equating the acceleration rate

\begin{equation}
	t^{\prime-1}_{\text{acc}} \equals \frac{\eta \, c^2 \, Z \, e \, B'}{E'},
	\label{equ:acceleration_efficiency}
\end{equation}

with the sum of synchrotron loss, photo-disintegration and photo-meson production rates, and the light-crossing time of the blob. Here, we choose $\eta=1$ for the acceleration efficiency\footnote{A case with $\eta = 0.1$ is discussed in \App{}~\ref{app:acceleration_efficiency}.}, and $B'$ is the magnetic field strength in the blob. Unless noted otherwise, the normalization of the nuclear injection spectrum is determined by a fixed ratio compared to the gamma-ray luminosity, namely the CR loading factor $f_{\text{CR}} = 10$. 

In order to set up the PDE system, automated techniques are used:
Starting with a database of nuclear isotopes~\cite{Mathematica160215} (white boxes in \figu{isotopes}), fast spontaneous nucleon or $\alpha$-particle emitters  (with a very conservative selection threshold $\tau_0 < 10^{-10} \, \mathrm{s}$) are integrated out from the beginning, which means that they are not explicitly considered but instead replaced by their daughters. Then, starting with the heaviest possible injection isotope iron-56, a recursive algorithm determines all possible photo-disintegration and decay (and also mixed) paths, \ie~the isotopes which will be dominantly populated. Such isotopes are marked in blue/green in the figure, and for these the coupled differential equation system is automatically set up and solved. The colored isotopes marked by crosses are potentially interesting (fast enough) beta emitters, which may contribute to neutrino production in the jet -- which we take into account. Included radiation processes for the nuclei are $\beta^\pm$ decays and spontaneous nucleon (and light nuclei) emissions~\cite{Mathematica160215}, photo-disintegration with TALYS~\cite{Koning:2007} ($A \ge 12$) and CRPropa2~\cite{Kampert:2012fi} ($A<12$), photo-meson production~\cite{Mucke:1999yb,Hummer:2010vx,Biehl:2017zlw}, photo-pair production~\cite{Blumenthal:1970nn}, synchrotron cooling and particle escape (depending on the CR escape assumption).\footnote{For the assumption of diffusive CR escape (see \sect{escape}) we include adiabatic cooling, leading to a decay of the density of charged particles after $t'_\mathrm{flare}$.}

As opposed to \Refs~\cite{Boncioli:2016lkt,Biehl:2017zlw}, we use a time-dependent computation for the fluences, \ie~we explicitly integrate the neutrino and CR fluxes instead of using the steady state approach. The solver integrates the blob system up to a few times $t'_{\rm flare}$, while the injection of nuclei is switched off at $t'_{\rm flare}$ (and the electrons are assumed to maintain the radiation field). While this approach is computationally more expensive than a steady-state solution and leads to similar results in the jet zone, it is more convenient for the modeling of the CR transport between the jet, BLR and DT zones (see \sect{fsrq}).

The magnetic field inside the blob plays a role in the cooling of charged CRs, mesons and muons through synchrotron radiation, in the escape rate of charged CRs if a diffusion mechanism is assumed (see \sect{escape}), and in the acceleration rate \equ{acceleration_efficiency}. The magnetic field is obtained through the relation
\begin{equation}
	u^\prime_B \equals \frac{\epsilon_B \, L'_\gamma \, t'_{\text{flare}}}{V'} \,\propto\, r^{'-2}_{\text{blob}},
	\label{equ:b}
\end{equation}
where $u^\prime_B \equals B^{\prime^2}/8\pi$ is the magnetic energy density and $\epsilon_B$ is the magnetic loading. We fix the value of $\epsilon_B$ for all sources in this study, with the brightest jet (introduced in \sect{sequence}) having a magnetic field strength of $5 \, \mathrm{G}$, following~\cite{Murase:2014foa}. Such constant value of the magnetic loading for all the sources, in this case $\epsilon_B=6.3\cdot10^{-3}$, leads to the magnetic field strength being proportional to $L_\gamma^{1/2}$ and inversely proportional to the size of the blob. In \App{}~\ref{app:acceleration_efficiency} we explore a different choice of magnetic field values, assuming the SEDs are generated via leptonic SSC models.

The blazar emission power in the source frame is represented by $L_{\gamma}$, the isotropic-equivalent integrated luminosity of the SED above 100 MeV. From the blazar sequence modeling, the SED and hence the comoving frame photon spectrum  are constructed phenomenologically as a function of $L_{\gamma}$ \cite{Murase:2014foa}.   The $\gamma-$ray emission power in the comoving frame of the blob, $L_{\gamma}^{\prime}$, is connected to the black-hole frame quantity via $L_{\gamma}=\Gamma^{4}L_{\gamma}^{\prime}$, where a factor $\Gamma^2$ comes from boosting the luminosity itself and a factor $\Gamma^2$ from the solid angle boost. Due to the fact that all the nuclear interactions (photo-meson production, photo-pair production and photo-disintegration) require a certain threshold energy, the relevant target photons are mainly the soft photons located in the first hump and the middle range of the SED.

We emphasize that the blazar SEDs are taken from observations, where we assumed that they emerge self-consistently from all participating radiation processes. However, it remains important to be checked if and under what conditions this assumption can hold. The feedback of hadronic photons to the SED is non-trivial and can be difficult to compute, even in the simplest proton-only case. As shown in Fig.4 of \cite{Gao:2016uld}, the feedback may show up in the X-ray range -- the saddle of the two humps. In this work, we aim to study which object classes can efficiently produce neutrinos or CRs. Therefore, we leave the fully self-consistent construction of SEDs, including nuclear interactions, to future works.

\section{Cascading of nuclei in the jet}
\label{sec:nuclear_regimes}

In this section, we discuss two distinctive scenarios where different physics takes place for nuclei in the jet: a low-luminosity prototype that is optically thin to photo-nuclear interactions, representative of what we call the ``nuclear survival'' regime, and a high-luminosity jet that is optically thick to photo-nuclear interactions at the maximum energy, representing the ``nuclear cascade'' regime. 

\subsection*{Nuclear survival regime}

\begin{figure*}[tbp!]
	\includegraphics[width=0.85\textwidth]{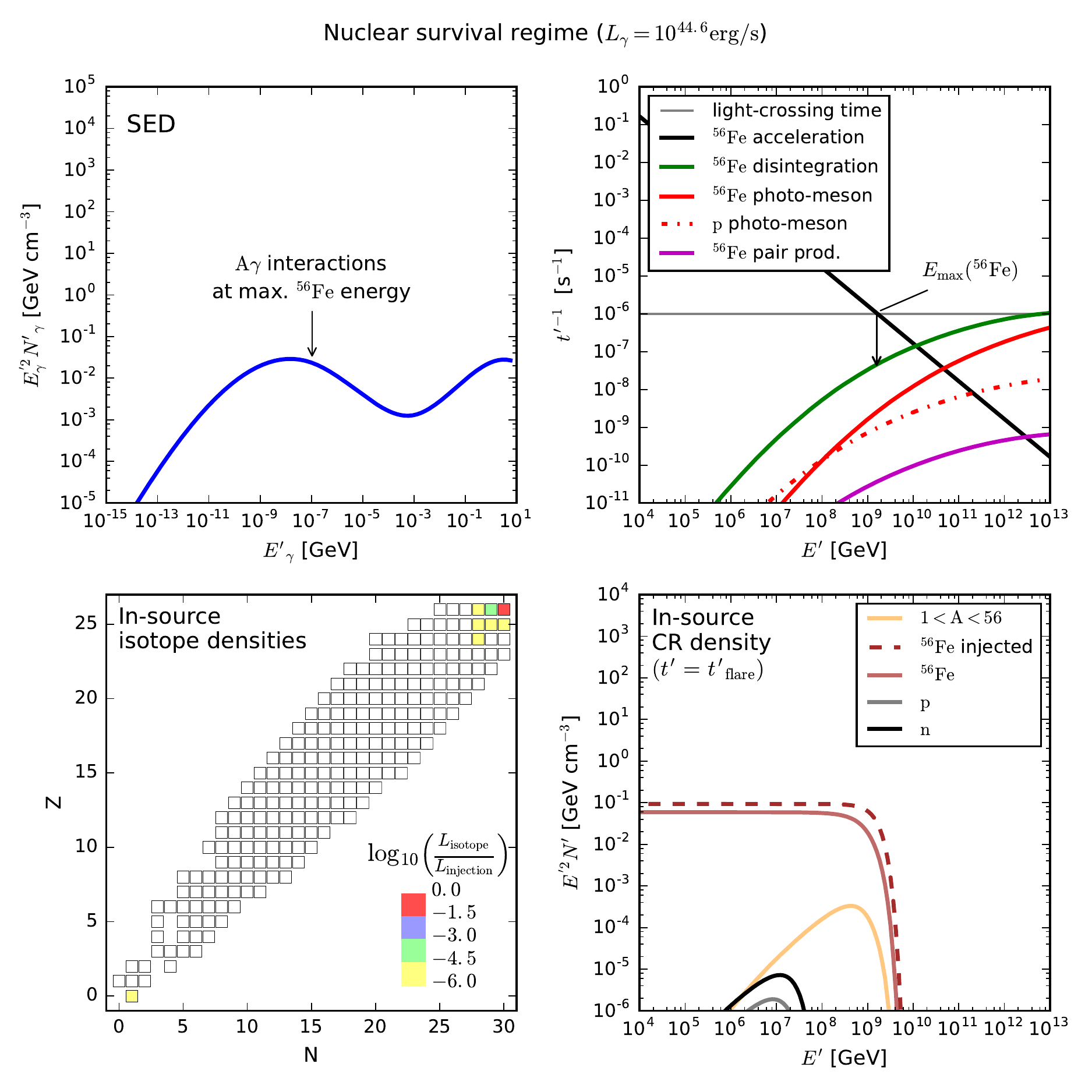}
	\caption{Example of the nuclear survival regime. Simulation of the injection of iron-56 in a jet with $L_\gamma=10^{44.6}$ erg/s. (\textit{Upper left}) time-independent SED; (\textit{upper right}) time scales (interaction rates) of relevant radiative processes and light-crossing time; (\textit{lower left}) energy densities of nuclear isotopes, normalized to the injection luminosity; (\textit{lower right}) spectra of baryons in the source after light-crossing time (given in the jet rest frame). The optical thickness to $A\gamma$ interactions $\tau$ at $E_{\rm max}$ is illustrated using black arrows.
}
	\label{fig:jet_44_5626}
\end{figure*} 

The example in \figu{jet_44_5626} illustrates the nuclear survival case for iron-56 injection, where $L_\gamma=10^{44.6}$ erg/s and the SED is shown in the upper-left panel. From the interaction rates in the upper-right panel, one can clearly see that the source is optically thin to disintegration at the maximum energy $E_\mathrm{max}^{\prime}\sim10^{9}$ GeV (which is determined by the light-crossing time) -- although disintegration, which strongly increases with energy, is the leading photo-nuclear process in the radiation zone. 
The initial nucleus interacts so rarely, that no significant cascading into lower mass nuclei can be observed (lower-left panel) and thus the primary nucleus survives. Since the maximum energy is determined by the light-crossing time, the primary spectrum in the source corresponds to the injection spectrum extending to the maximum injection energy and without spectral breaks (see lower-right panel). The small offset between the injection spectrum and the in-source density of iron-56 is due to adiabatic cooling (see \sect{escape}).

\subsection*{Nuclear cascade regime}

\begin{figure*}[tbp!]
	\includegraphics[width=0.85\textwidth]{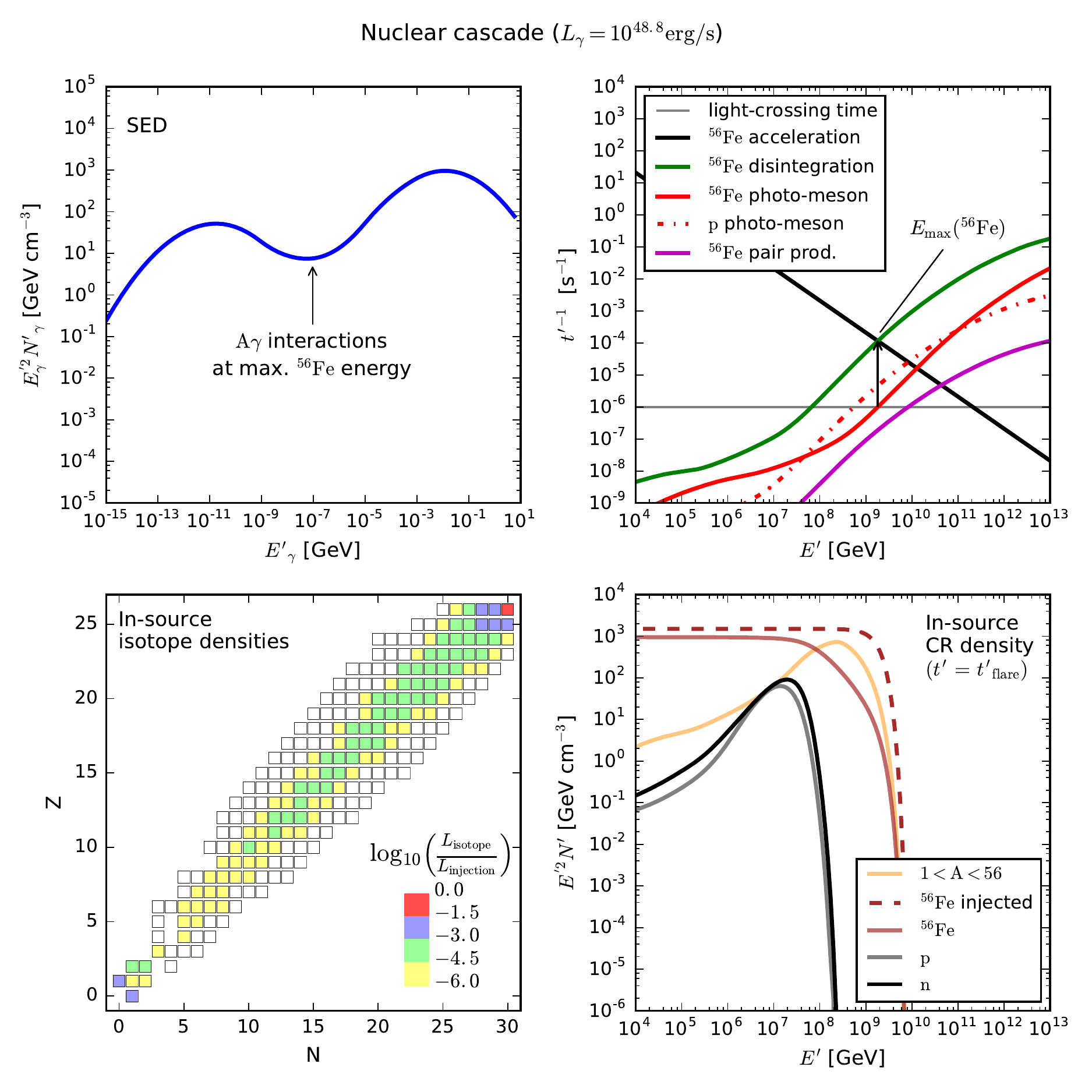}
	\caption{Example of the nuclear cascade regime (jet with $L_\gamma=10^{48.76}$ erg/s), for a pure iron-56 injection. See caption of \figu{jet_44_5626} for details.
}
	\label{fig:jet_49_5626}
\end{figure*}

The nuclear cascade regime is illustrated in \figu{jet_49_5626} by a high-luminosity jet with $L_{\gamma}^{\prime}=10^{48.8} \rm{erg/s}$ for iron-56 injection, where the corresponding SED is shown in the upper-left panel. Note that only interactions with the jet radiation are considered, without additional components from BLR or DT. The maximum energy is limited by photo-disintegration, and the source is optically thick ($\tau \sim 100$) to this process (see upper-right panel), resulting in efficient disintegration of the primary nucleus into lower mass nuclei (see lower-left panel). Significant fractions of energy are transfered into densities of neighboring nuclei and light fragments, such as nucleons and $\alpha$ particles. As opposed to high target density cases in gamma-ray bursts (GRBs)~\cite{Biehl:2017zlw}, the disintegration rate strongly decreases with energy. The in-source density at the highest energies (lower-right panel) is dominated by secondary nuclei from photo-disintegration reprocessing the injected isotopes into a lighter mass composition. This is characteristic for the high-energy CRs in the nuclear cascade regime, whereas the primary spectrum at lower energies is not notably modified. 

Regarding the dominant radiation process (upper-right panel), we find that disintegration typically dominates over photo-meson production for the primaries. It has been demonstrated in \Ref~\cite{Anchordoqui:2007tn}, as reproduced in this work, that photo-meson production can dominate over photo-disintegration for special SEDs in certain energy ranges. We observe this behavior in certain rare SEDs at the most extreme luminosities, which we however do not consider to be representative. Note that in both \Ref~\cite{Anchordoqui:2007tn} and this work, a superposition model for photo-meson production has been used whereas the realistic photo-meson cross sections will be somewhat smaller and need further study; see discussions in \Ref~\cite{Boncioli:2016lkt,Biehl:2017zlw}.

\subsection*{Parameter space study}
\begin{figure*}[t!]
	\includegraphics[width=\textwidth]{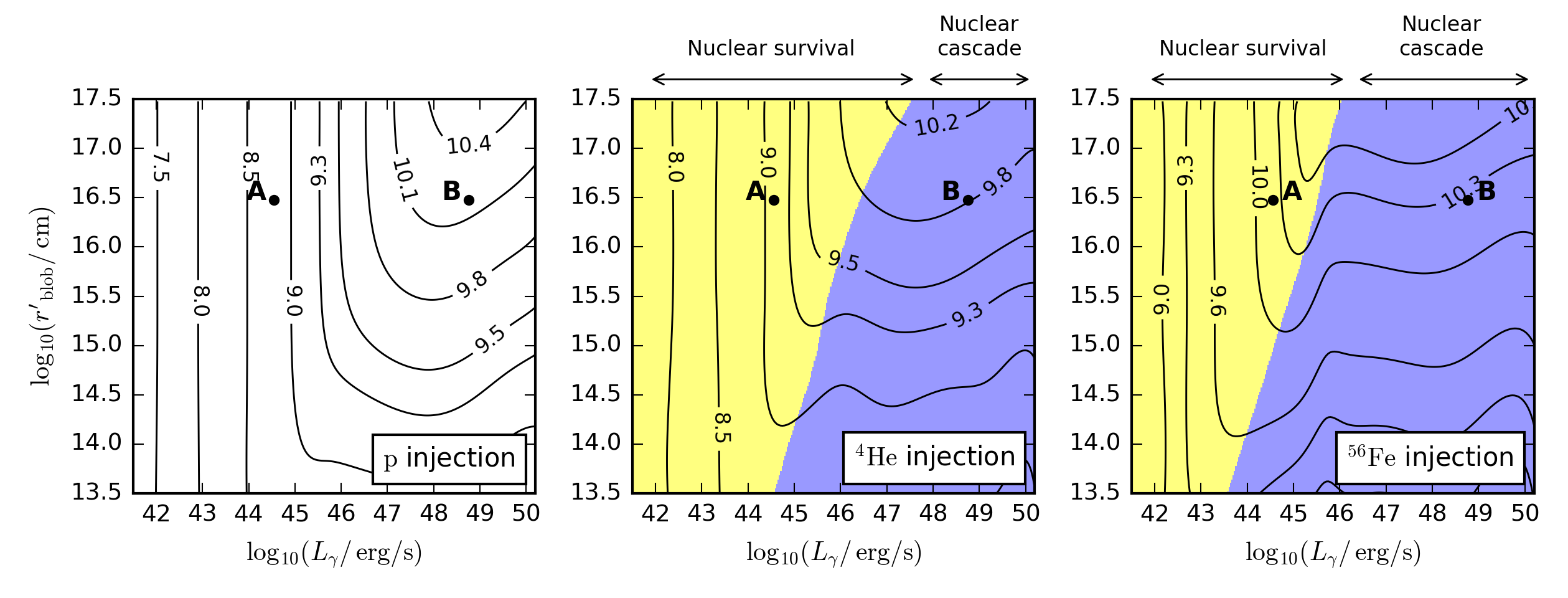}
	\caption{Maximum injection energy and nuclear regimes obtained for a range of jet luminosity and blob size values, for the injection of three different isotopes. The contour labels indicate the maximum injection energy in the black-hole frame, $\text{log}_{10}(E_{\text{max}}/ \text{GeV})$ (see \sect{jet_model}). In the center and right-hand panels, the two colors represent the two nuclear regimes. The points marked \textit{A} and \textit{B} represent the position in parameter space of the nuclear survival prototype (\protect{\figu{jet_44_5626}}) and the nuclear cascade prototype (\protect{\figu{jet_49_5626}}), respectively.}
	\label{fig:parameter_scan}
\end{figure*}

By scanning the parameter space of $r_\mathrm{blob}^{\prime}$ and $L_{\gamma}$, we find that the two nuclear regimes, defined by the optical thickness to photo-nuclear interactions at the highest energy, can occur over a wide range of parameters. We show this result in \figu{parameter_scan} for three different injection isotopes: protons, helium-4 and iron-56. Note that in this figure, the shape of the SED is assumed to follow the blazar sequence as a function of $L_\gamma$ as we will introduce later in \Sec~\ref{sec:sequence} (but without the external radiation fields), whereas $r_\mathrm{blob}^{\prime}$ only affects the photon density normalization.  From \figu{parameter_scan} we see that by varying $L_{\gamma}$ along a fixed line of $r_\mathrm{blob}^{\prime}=3.0\times10^{16}$ cm, corresponding to our prototypes (points $A$ and $B$), one can effectively control the level of disintegration and, as we will see later, the balance between a more efficient neutrino source versus a more efficient UHECR source. A similar effect can be achieved by changing $r_\mathrm{blob}^{\prime}$ for fixed $L_\gamma$. As we will later associate $L_\gamma$ with certain object classes of the blazar sequence, it is important to remember that the splitting point between nuclear survival and nuclear cascade depends somewhat on the blob size. Comparing the center and right-hand panels of \figu{parameter_scan} we see that iron-56 can populate a nuclear cascade at lower jet luminosities compared to helium-4, since iron-56 has a higher photo-disintegration rate.

Figure~\ref{fig:parameter_scan} contains contours of the the maximum energy, and it can be clearly seen that larger accelerators tend to allow for higher acceleration energies. For the case of protons (left panel), these are determined by the available time for acceleration in the case of lowest luminosities, while at the highest luminosities, photo-meson interactions are responsible for limiting the maximum energy. For the case of nuclei, photo-disintegration limits the maximum energy at the highest luminosities and determines the transition to the nuclear cascade regime. For low luminosities the maximum energy does not depend on the size of the blob, neither for protons nor for the nuclei. That is because the dynamical time scale is directly proportional to the blob size, while the acceleration rate is proportional to the magnetic field, which scales inversely with the blob size. Above a certain luminosity the shape of the contours is more complicated due to the interplay between the acceleration and photo-disintegration rates. The central and right-hand side panels of \figu{parameter_scan} reveal that the maximum energy is obtained in a narrow region where the transition between the two regimes occurs. In this region the acceleration is strong enough to reach high energies, while energy losses from photo-disintegration are not too restrictive. The small features in the contours originate from the variation of SEDs along the blazar sequence.

\section{Neutrino production}
\label{sec:neutrinos}

\begin{figure*}[t!]
	\includegraphics[width=0.9\textwidth]{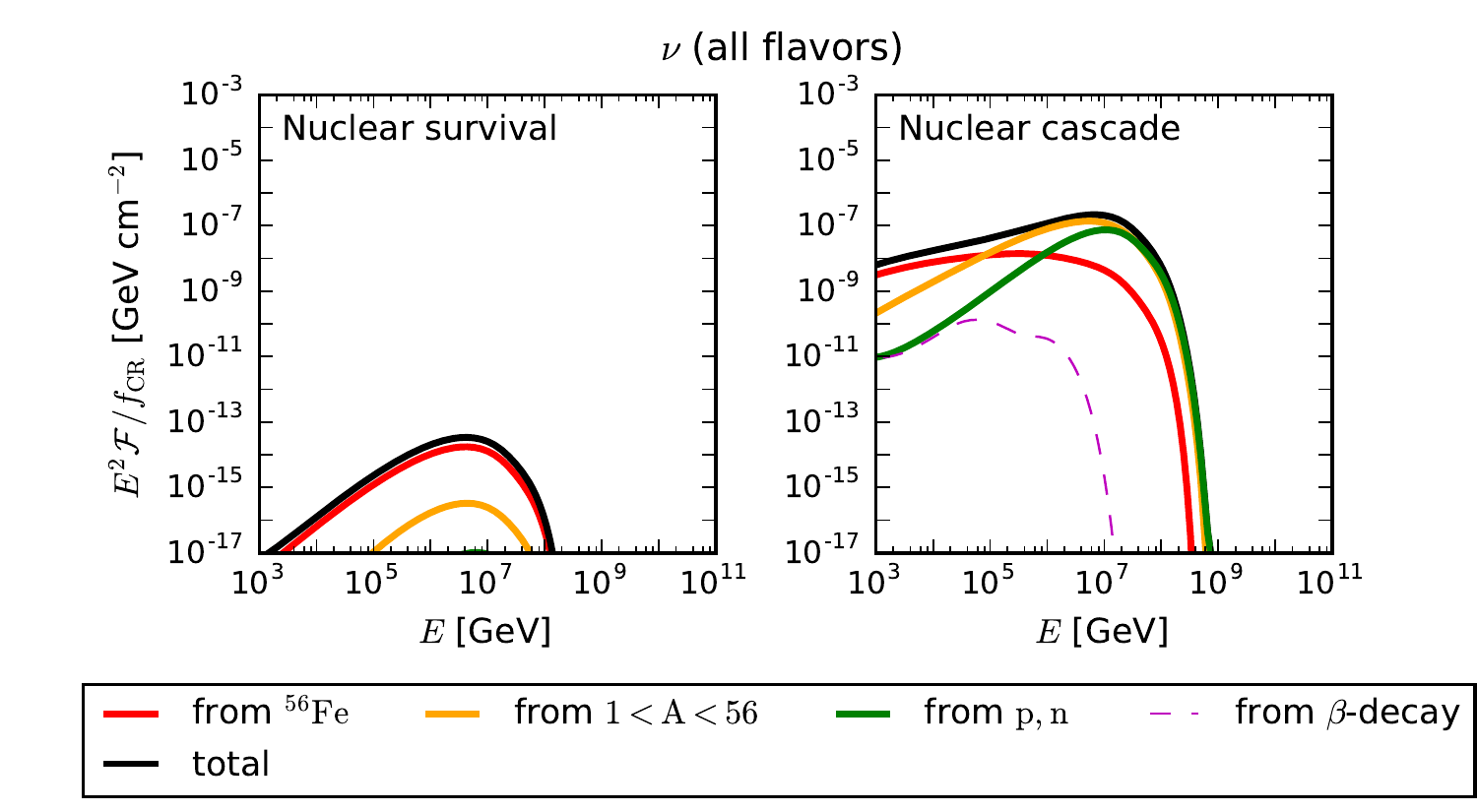}
	\caption{Ejected all-flavor neutrino fluence for the nuclear survival and nuclear cascade examples, for a pure iron-56 injection composition, where the contributions from interactions of the primary nucleus, the secondary nuclei, the nucleons, and $\beta$ decay are shown separately. The fluences here (and elsewhere in this study) are shown in the observer's frame, considering a source redshift of $z=1$; the fluences are also divided by the CR loading which is an arbitrary parameter in this study, \ie~they need to be multiplied with $f_{\mathrm{CR}}$ to obtain the absolute values.}
	\label{fig:neutrino_spectra}
\end{figure*}

The ejected neutrino fluence for both the nuclear survival and cascade regimes is shown in \figu{neutrino_spectra}, where the contributions from interactions of the primary, the secondary nuclei, the nucleons, and $\beta$ decay are shown separately (where the secondary nuclei and nucleons are produced in the nuclear cascade). In the nuclear survival regime neutrinos are produced off the injection isotope, with very small contributions from secondary nuclei and nucleons. On the other hand, for sources with a nuclear cascade, most of the neutrinos at high energies are produced off secondary isotopes and nucleons. 

Contrary to the GRB case, discussed in \cite{Biehl:2017zlw}, where three different regimes were identified\footnote{Nuclear survival case, where neutrino production is dominated by photo-meson production off the primary, nuclear cascade case, where neutrino production is dominated by photo-meson production off the secondary nuclei produced by disintegration, and optically thick case, where neutrino production is dominated by photo-meson production off nucleons.}, we only find two distinctive scenarios for blazars. The main difference is that for GRBs the interaction rate is about constant beyond the break in the SED, whereas it strongly increases for blazars in the relevant energy range (see \Figs~\ref{fig:jet_44_5626} and \ref{fig:jet_49_5626}, upper right-hand panels). While the nuclei in GRBs continue to disintegrate efficiently along the cascade, in blazars disintegration ceases much faster because the disintegration rate quickly drops with decreasing energy. We therefore do not find a case where most energy is dumped into nucleons, that would be comparable to the optically thick regime of the GRB models \cite{Biehl:2017zlw}. Rather, neutrino production in the nuclear cascade regime of blazars is typically dominated by the secondary nuclei produced in the nuclear cascade. This again highlights the importance of better photo-meson production models off nuclei pointed out in \Ref~\cite{Biehl:2017zlw}, as our results are based on a simple superposition model where the total interaction and inclusive pion production cross sections scale with $A$. 

\begin{figure*}[t!]
	\includegraphics[width=0.9\textwidth]{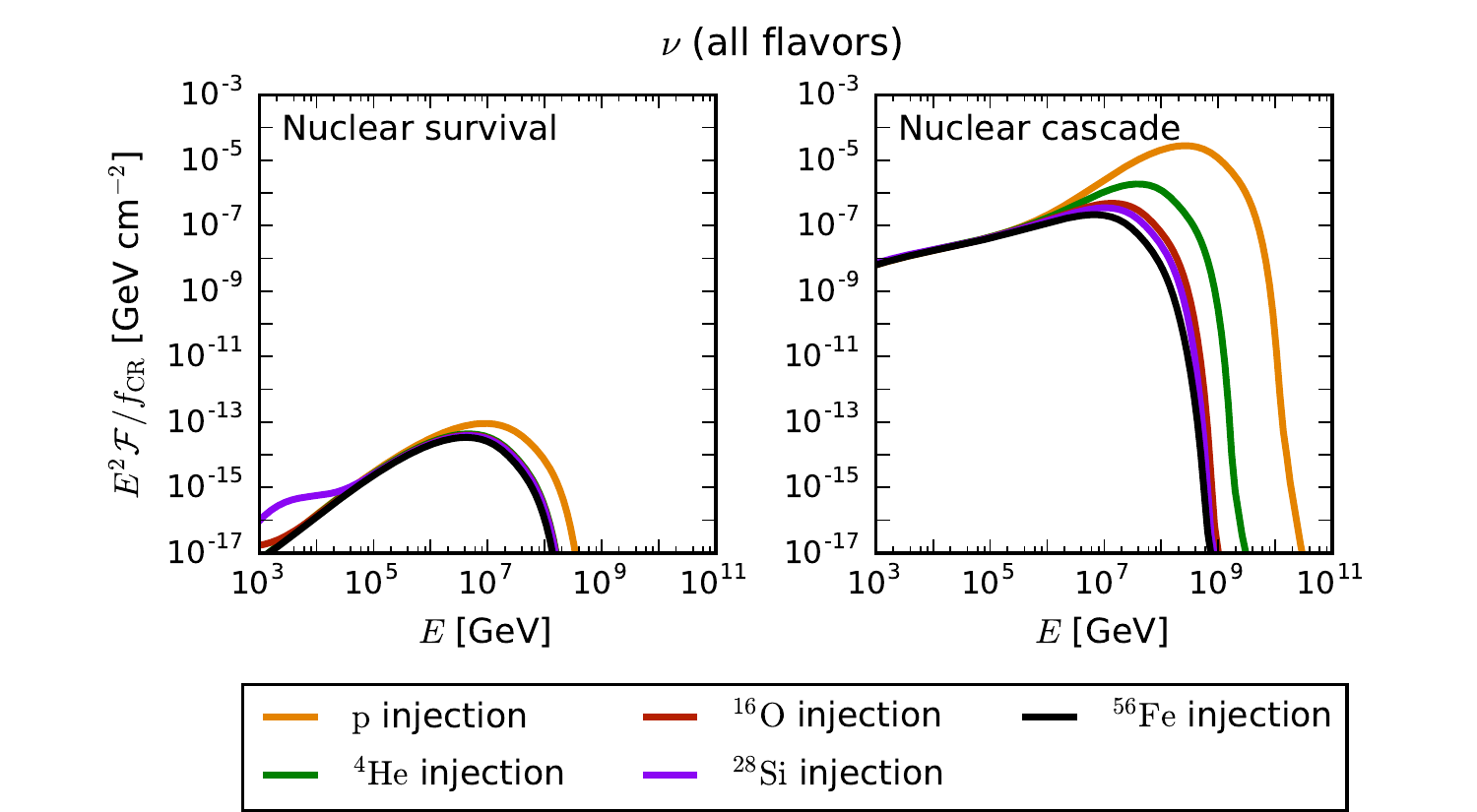}
	\caption{All-flavor neutrino fluence for the nuclear survival and nuclear cascade examples for different injected isotopes, shown in the observer's frame for redshift $z=1$. 
}
	\label{fig:nu_composition}
\end{figure*}

The dependence of the neutrino spectrum on the injection composition is shown in \figu{nu_composition}. In the nuclear survival case, the maximum energy per nucleon scales $\propto Z/A \simeq 1/2$ for most isotopes (except nucleons), which means that we observe a weak dependence on the composition. In the nuclear cascade regime, the neutrino spectrum basically follows the maximum energy per nucleon as well, which however depends in a non-trivial way on the competition of the disintegration, photo-meson and acceleration time scales. In fact, from \figu{parameter_scan}, one can see that the maximum primary energy $E_{\mathrm{max}}$ for point~B varies only a factor of a few with injection composition, which means that the neutrino energy roughly scales $\propto E_{\mathrm{max}}/A$ following the energy/nucleon. Consequently, the neutrino spectrum extends to higher energies for lighter injection compositions.
We therefore find that loading (high-luminosity) blazars with nuclei helps reduce the tension with the non-observation of blazar neutrinos by IceCube both in terms of spectrum and normalization.

\section{Cosmic-ray escape}
\label{sec:escape}

We have previously discussed CR densities inside the source, which are relevant for neutrino production in the jet. In order to address CR injection into the intergalactic space, one has to specify how CRs escape from the source. 

The escape of particles from the source environment on the macroscopic scale is typically modeled via diffusion and advection processes. In order to fully address this effect, the detailed profile of the plasma, such as the geometry, plasma wave and magnetic field configurations are required to compute the diffusion and convection coefficients on a spatially resolved grid \cite{Chen:2016ram}. However, in one-zone calculations, a global energy-independent escape rate is frequently assumed \cite{Diltz:2015kha,Petropoulou:2015upa,Gao:2016uld}, as a fraction of the free-streaming rate. In this work, while keeping the simplicity and efficiency of the one-zone calculation, we use a more refined approach: We distinguish between escape by diffusion and by advection, two scenarios that correspond respectively to the most conservative and most aggressive escape assumptions. We expect that any realistic escape assumption will be in between these two scenarios. In either case, the maximum particle energy is computed self-consistently from equating acceleration, escape and energy loss rates.

{\bf Escape by diffusion/direct escape.} Escape by diffusion of charged CRs can be described by an effective escape rate $t'^{-1}_{\mathrm{esc}} = D' \,  t'^{-2}_{\mathrm{flare}}/c^2$ (from $r'_{\mathrm{blob}}= \sqrt{ D' \, t'_{\mathrm{esc}}}$), where $D'$ is the (spatial) diffusion coefficient.\footnote{Here it is implied that the coherence length of the magnetic field is large enough; effects of the coherence length can slightly alter the shape of the spectrum, but do not affect the conclusion of hard spectra and that all particles escape if the Larmor radius exceeds the size of the region; see e.g.\ \Ref~\cite{Kimura:2017ubz}.} For Bohm diffusion, which may be the most extreme assumption here, $D' \simeq c R'_L$, where $R'_L \propto E'$ is the Larmor radius, which leads to hard escape spectra. If the Larmor radius can reach the size of the blob (implying that the maximum energy is determined by the flaring timescale), all particles that do not interact will escape. Note that this scenario conservatively implies magnetic confinement of charged nuclei, while neutrons can escape over the free-streaming timescale. We therefore include adiabatic cooling of the charged species (which is expected if the blob expands) allowing the particle densities to decay after $t'_{\mathrm{flare}}$.

\begin{figure*}[tb!]
	\includegraphics[width=0.9\textwidth]{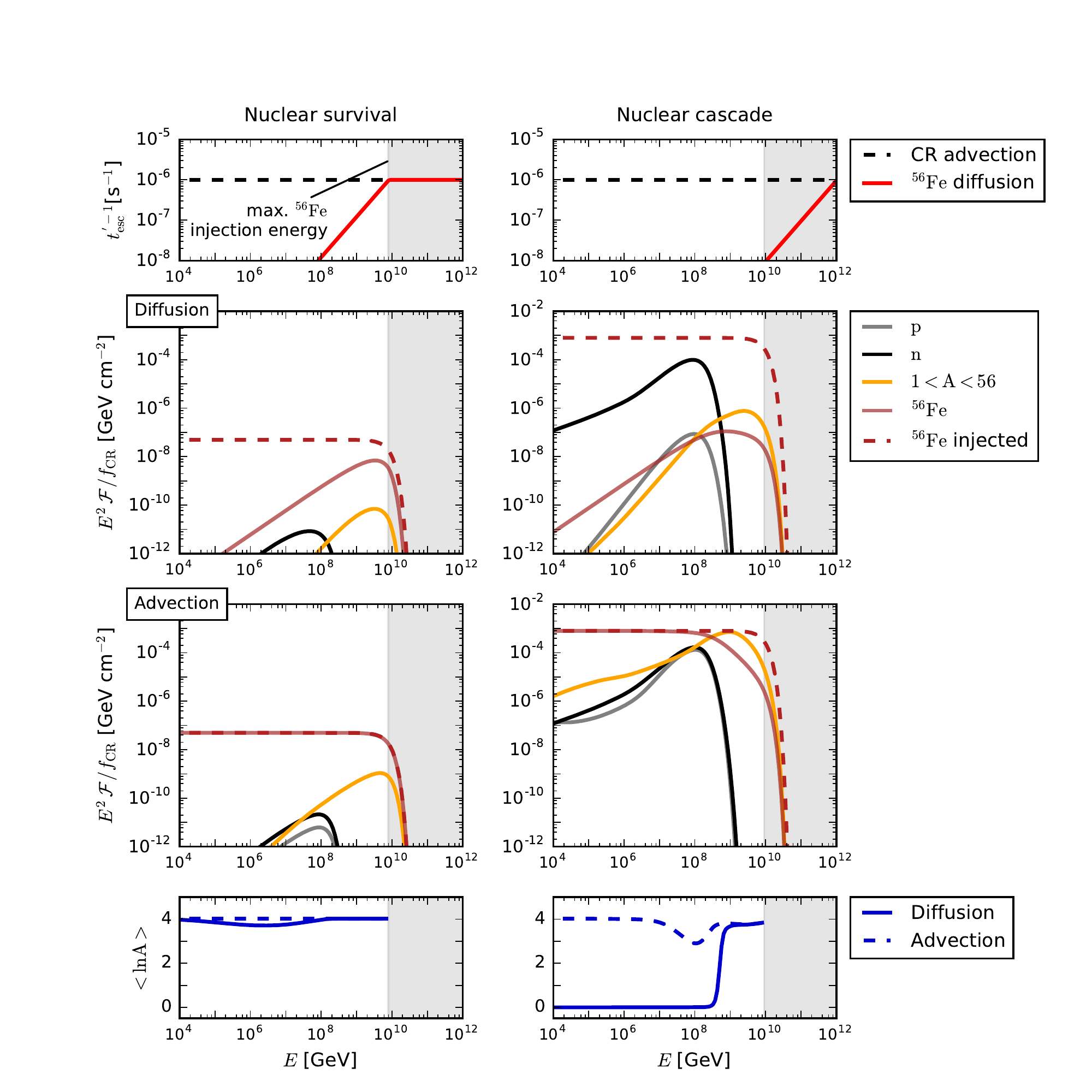}
	\caption{Cosmic ray escape rates, ejected CR spectra and average CR composition for the nuclear survival and nuclear cascade
examples. \textit{Top row:} Escape rates (in the blob rest frame) for a diffusion (solid) and advection (dashed) escape mechanism. \textit{Second row:} Ejected spectra for diffusive CR escape, given in the observer’s frame considering a redshift $z=1$. \textit{Third row:} Ejected spectra for advective CR escape. \textit{Bottom row:} Average ejected CR composition for both escape mechanisms. The energies on the horizontal axes are in the observer’s frame. Note that no interactions in the propagation are included.}
	\label{fig:cr_esc_prototypes}
\end{figure*}

In the upper panels of \figu{cr_esc_prototypes} we show the rates for the diffusion-dominated escape as solid curves. In the nuclear survival case, the maximum energy is limited by the light-crossing time, which implies that the Larmor radius at the highest energy can reach the size of the blob; however the escape rate cannot exceed the free-streaming escape rate.  Up to the maximum energy, the escape rate grows with the Larmor radius $\sim E$; it therefore acts as a ``high-pass filter'' on the ejected CRs and results in hard emission spectral indices (see solid curve in the left-hand panel in the 2nd row). In the nuclear cascade case (see upper right-hand panel), the maximum energy is limited by photo-disintegration, which means that the Larmor radius cannot reach the size of the blob at the maximum energy, and only a fraction of the charged particles can escape. Neutrons, however, which are abundantly produced in the nuclear cascade, can freely escape. The corresponding escape spectra are shown as solid curves in the right-hand panel in the 2nd row.
 
A similar escape mechanism has been assumed by \Refs~\cite{Baerwald:2013pu,Globus:2014fka,Zhang:2017hom} for GRBs and Tidal Disruption Events respectively; for example, the assumption that only a fraction close to the border of the region proportional to the Larmor radius can escape  (``direct escape'') leads mathematically to the same results. Another motivation for the preference of harder spectra comes from a recent result by the Auger Observatory, which suggests that source spectra are preferentially hard \cite{Aab:2016zth} when fitting a generic class of UHECR accelerators. 

{\bf Escape by advection.} Cosmic rays transported along the jet effectively escape if the magnetic fields and densities decay quickly enough and cooling is inefficient. We assume free-streaming escape as the most aggressive escape assumption in this case. The corresponding escape rates are shown as dashed curves in the upper panels of \figu{cr_esc_prototypes}.

Note that our advective escape assumption is similar to \Ref~\cite{Murase:2014foa} in the nuclear survival case, where the  effective escape has been described by $t'^{-1}_{\mathrm{esc}} = f_{\mathrm{esc}} \,  t'^{-1}_{\mathrm{flare}}$ with the escape fraction $f_{\mathrm{esc}}=(1-\min[1,t'_{\mathrm{flare}}/t'_{\mathrm{cool}}])$ and $t'_{\mathrm{cool}}$ the dominant cooling timescale. However, compared to \Ref~\cite{Murase:2014foa}, the effect of photo-hadronic cooling is already included in our self-consistent nuclear cascade computation (the in-source spectra are suppressed if \eg\ photo-disintegration dominates), which is why we do not take into account an additional cooling-driven suppression (which is relevant in the nuclear cascade case). Note that it is a prerequisite for advective escape that the CRs must not cool over a timescale which is much longer than $t'_{\mathrm{flare}}$ if the interactions in the BLR are to be taken into account. In particular, any adiabatic expansion of the blob would suppress this escape component, which may be evaded if the jet is collimated. Consequently, we do not include adiabatic cooling in this scenario, and after $t'_{\mathrm{flare}}$ densities in the source simply decay by escape and interactions.

The 2nd and 3rd rows in \figu{cr_esc_prototypes} compare the effect of the two escape hypotheses on the ejected CR spectra. The total escape spectra in the advective case resemble the $E^{-2}$ injection spectrum over a wide range of energies since the escape rate is flat in energy, \ie~they are generally softer than in the diffusive case. Note that in the nuclear survival case (left-hand side panels), near the spectral cutoff, the fluence is higher for the advective escape assumption simply because of the absence of an adiabatic cooling term. The diffusion assumption, on the other hand, hardens the spectra of charged particles by one power (2nd row). Since neutrons are not magnetically confined, their spectra are identical for both escape assumptions.
As mentioned previously, in reality we expect the escape spectrum to lie in between these two extreme cases.

The ejected mass composition in the advection scenario is orthogonal to what is expected from UHECR observations (see bottom right-hand panel). Since in this study we are mostly interested in UHECRs, we conclude that in the nuclear cascade case a diffusive escape mechanism is potentially more effective, at the highest energies, in producing the right trend of the ejection composition as a function of energy within the source itself; whereas in the nuclear survival case, it allows for hard ejection spectra compatible with the Auger fit~\cite{Aab:2016zth}.

\section{Model for HL-FSRQs}
\label{sec:fsrq}

For FSRQs, radiation from outside the jet is observed as well, such as the thermal radiation from the dust torus (DT) and the line emissions from the so-called broad line region (BLR). In our model, only HL-FSRQs may have BLR and DT regions large enough to physically contain the blob (see middle panel of \figu{model}). The photons from the accretion disk can be back-scattered in the BLR and have physical importance for interactions in the blob. At the same time, messengers produced in the blob may interact in the BLR and the DT, producing neutrinos and reducing the total CR output.

For the parameterization of the additional zones, we follow the model from \Ref~\cite{Murase:2014foa}; see our \figu{model} (lower panel). Note that the classification of FSRQs into high- and low-luminosity is performed in order to identify sources which require additional radiation zones that impact secondary production. The classes of FSRQs with lower $L_{\gamma}$ (\ie~blazars where broad lines are present but have negligible impact on secondary production), are treated with the jet model only, equal to BL Lacs with similar $L_{\gamma}$. In greater detail, the additional ingredients for the HL-FSRQs are:
\begin{enumerate}
 \item External radiation from the accretion disk is reprocessed and isotropized in the BLR and DT and boosted into the blob, where it contributes to the target photon spectrum in addition to the non-thermal radiation (as illustrated in the middle panel of \figu{model}).
\item
  Cosmic rays escaping from the jet zone are boosted into the black-hole frame and re-processed in the external photon fields as they travel through the BLR, where again they may interact or escape. Those that escape propagate through the DT region, where they encounter thermal radiation from the warm dust. In both zones, the free-streaming time scale, which gives the escape rate of all CR species, corresponds to the size of the region.
\item In the BLR and DT, additional photo-meson production may occur. The neutrinos produced contribute to the total neutrino fluence of the source.
\end{enumerate}

In order to illustrate the effect of the external radiation fields, we use as an example an FSRQ with luminosity $L_\gamma=10^{48.8}$ erg/s, with the same jet properties as introduced in \Sec~\ref{sec:jet_model} (see \figu{jet_49_5626}). While the same magnetic field strength is consided in the jet, we assume the magnetic field strength in the BLR and DT to be much smaller, \ie~$B^{\prime}\ll1$ mG, a valid assumption if these regions consist of gas and clouds analogous to supernova remnants. Consequently, we neglect synchrotron emission and magnetic confinement in the BLR and DT.

\begin{figure*}[tb!]
	\includegraphics[width=\textwidth]{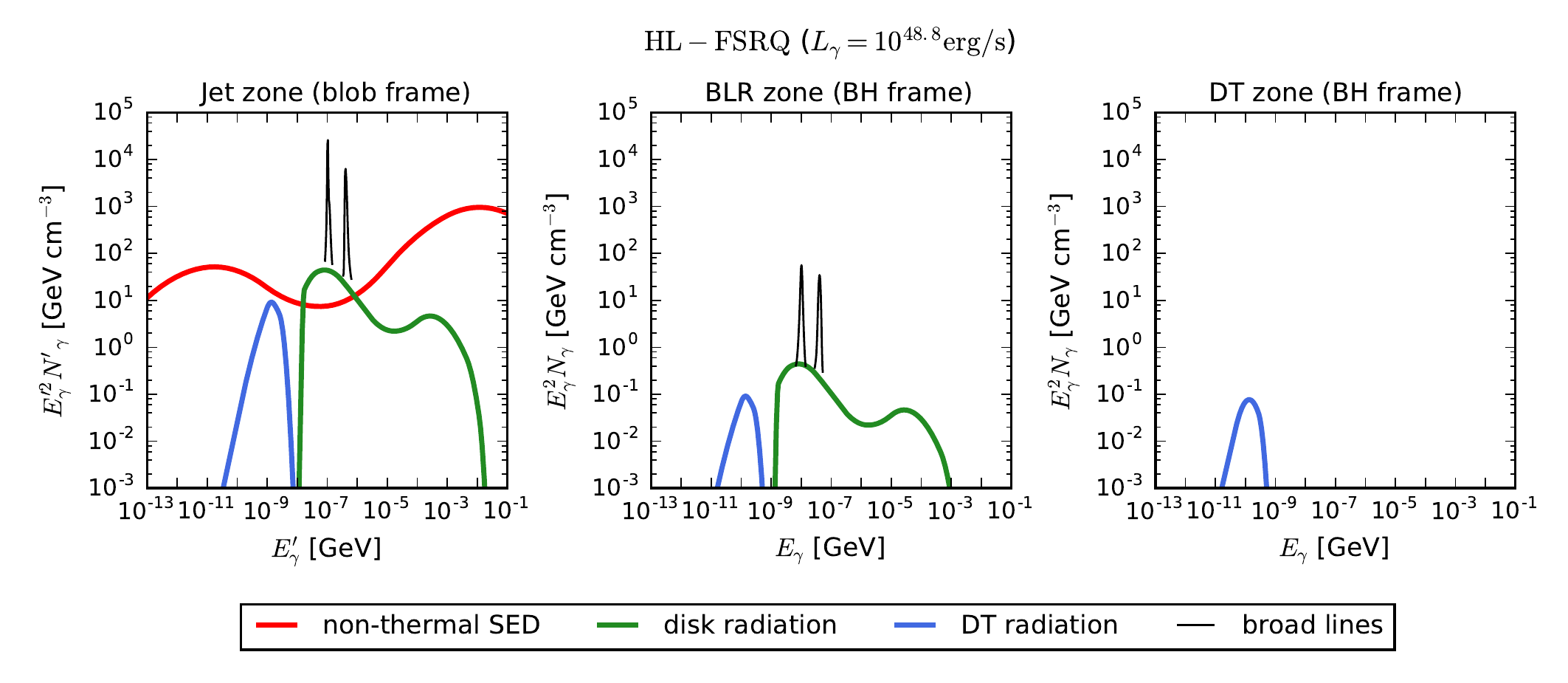}\vspace*{-5mm}
	\caption{Target photon densities in the three zones of a HL-FSRQ with $L_\gamma=10^{48.8} \text{erg/s}$ (with jet parameters as in \protect{\tabl{sequence}}), adopted from~\protect{\cite{Murase:2014foa}}. Besides the non-thermal radiation also present in BL Lacs \textit{(red)}, there are external components consisting of reprocessed accretion disk radiation: a scattered component spanning from the visible to the X-ray \textit{(green)}, the H I and He II Ly$\alpha$ emission lines \textit{(black)}, and an infrared thermal distribution from a dust torus at $T=500$ K surrounding the BLR \textit{(blue)}. See main text for details. Note that the photon fields in each of the blazar zones are given in their own rest frame.}
	\label{fig:zones_sed}
\end{figure*}

We illustrate the target photon densities in the three zones in \figu{zones_sed}, each in the respective reference frame. Unlike BL Lacs and lower-luminosity FSRQs, the jet zone receives additional target photons from the external radiation produced in the BLR and the DT. Starting with the accretion disk radiation, we assume that a fraction $\tau_{\text{sc}}$ is isotropized in the BLR through Thomson scattering, and that this component is Lorentz-boosted into the blob. This fraction is given by the optical thickness of the BLR to Thomson scattering, assumed to be $\tau_{\text{sc}} = 0.01$~\cite{Blandford:1995yf,Tavecchio:2012um}. In the middle panel of \figu{zones_sed} we show in green the density spectrum of photons from the disk that are scattered in the BLR, based on \Ref~\cite{Elvis:1994}. The total energy density of scattered photons $u_{\text{disk}}$ corresponds to the integral of this spectrum and is normalized according to the total disk luminosity using
\begin{equation}
	u_{\text{disk}} \equals \frac{\tau_{\text{sc}}}{4\pi\, r_{\text{BLR}}^2\, c} \, L_{\text{disk}} \, ,
	\label{equ:disk_photons}
\end{equation}
where $L_{\text{disk}}$ is the disk luminosity, given by the empirical relation discussed in \sect{methods}. Like all external components, this radiation is assumed to be isotropic in the BLR. In the jet rest frame, external photons appear boosted due to the relativistic motion of the blob, thus contributing to the target photon spectrum for CR interactions in the jet zone. The photon energy in the jet rest frame is then given by $E'_\gamma=\Gamma E_\gamma$, and the energy density receives a factor $\Gamma^2$ compared to the BLR (see left panel of \figu{zones_sed}).
We further consider the H I and He II Ly$\alpha$ lines emitted by the BLR gas molecules, which are excited by the disk radiation (peaks in \figu{zones_sed}). In our simplified approach, the photon density of the hydrogen line in the BLR is given by
\begin{equation}
	u_{\text{H I}} \equals \frac{f_{\text{cov, H}}}{4\pi\, r_{\text{BLR}}^2\, c} \, L_{\text{disk}},
	\label{equ:bl_photons}
\end{equation}
where $f_{\text{cov, H}}$ is the hydrogen covering factor of the BLR, assumed to be $0.1$ \cite{Ghisellini:2008zp,Liu:2006ja}, which represents the fraction of the disk luminosity re-emitted as the hydrogen broad line. The helium line density is calculated similarly but using a covering factor twice as small, so that $u_{\text{H I}}=2u_{\text{He II}}$~\cite{Tavecchio:2008vq}.
Finally, the thermal photon density from the DT is shown as blue curves in \figu{zones_sed}. This is a black-body distribution of temperature 500 K (in accordance with typical DT temperatures of 100-1000 K~\cite{Cleary:2006pe,Malmrose:2011ne}). Besides being present in the DT zone itself, this component also contributes to the target photon spectrum in the BLR as well as the jet, since the DT is assumed to surround these zones (see lower panel of \figu{model}). The total IR photon density is given in the black-hole frame by
\begin{equation}
	u_{\text{IR}} \equals \frac{f_{\text{cov, DT}}}{4\pi\, r_{\text{DT}}^2\, c} \, L_{\text{disk}},
	\label{equ:dt_photons}
\end{equation}
where $f_{\text{cov, DT}}=0.5$~\cite{Murase:2014foa} is the value assumed for the DT covering factor.

Note that in this model, the photon density of the external fields (disk, broad lines and IR radiation from the DT) does not vary for sources with different disk luminosity or BLR size, since
\begin{equation}
	u_{\text{disk; BL; IR}} \,\propto\, \frac{L_{\text{disk}}}{r_{\text{BLR}}^2} \simeq \text{const},
	\label{equ:photon_density_scaling}
\end{equation}
because of the scaling expressed in \equ{disk_luminosity}.

\begin{figure*}[tb!]
	\includegraphics[width=\textwidth]{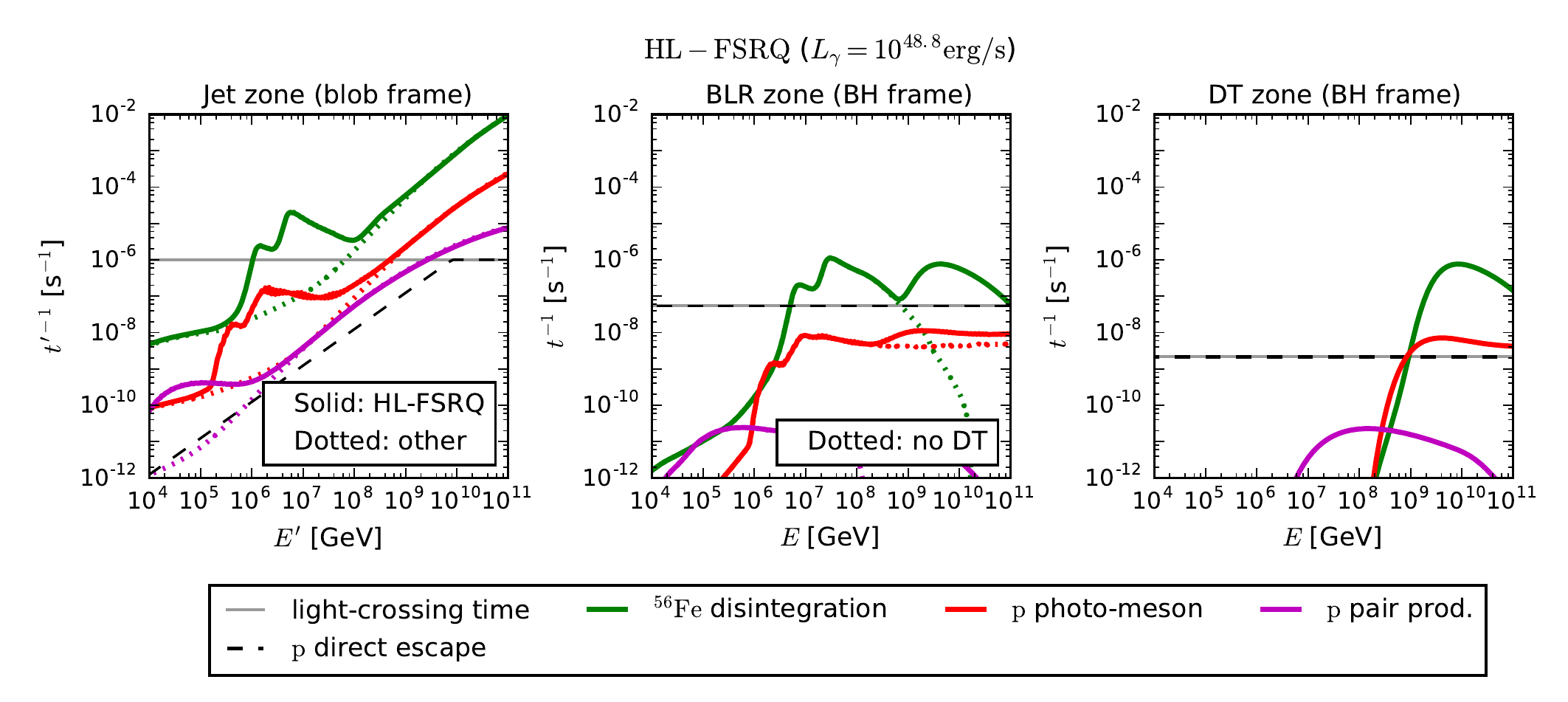}
	\caption{Interaction rates in the three radiation zones of the FSRQ model, in the rest frame of the zone, for the same source as in \figu{zones_sed}. Note that the rates in the BLR and DT zones (middle and right-hand panels) are independent of the size of the region (see \protect\equ{photon_density_scaling}). By comparing the solid and dotted lines, we see the effect of the external radiation fields in the inner zones of the model.}
	\label{fig:zones_rates}
\end{figure*}

The impact of external radiation fields on the interaction rates is shown in \figu{zones_rates}. While photo-disintegration (of iron-56 in this case) increases in the jet zone and the source becomes optically thick to photo-disintegration above PeV energies, protons are not significantly affected, because the source is still optically thin to photo-meson production. The nuclei that escape from the jet zone into the BLR and DT then undergo additional disintegration at the highest energies, reducing the efficiency of HL-FSRQs as UHECR sources. In cases where the BLR radiation is present but the DT is too small, the additional transport through the BLR would not significantly affect the UHECRs (dashed curves in central panel).

\begin{figure*}[tb!]
	\includegraphics[width=0.9\textwidth]{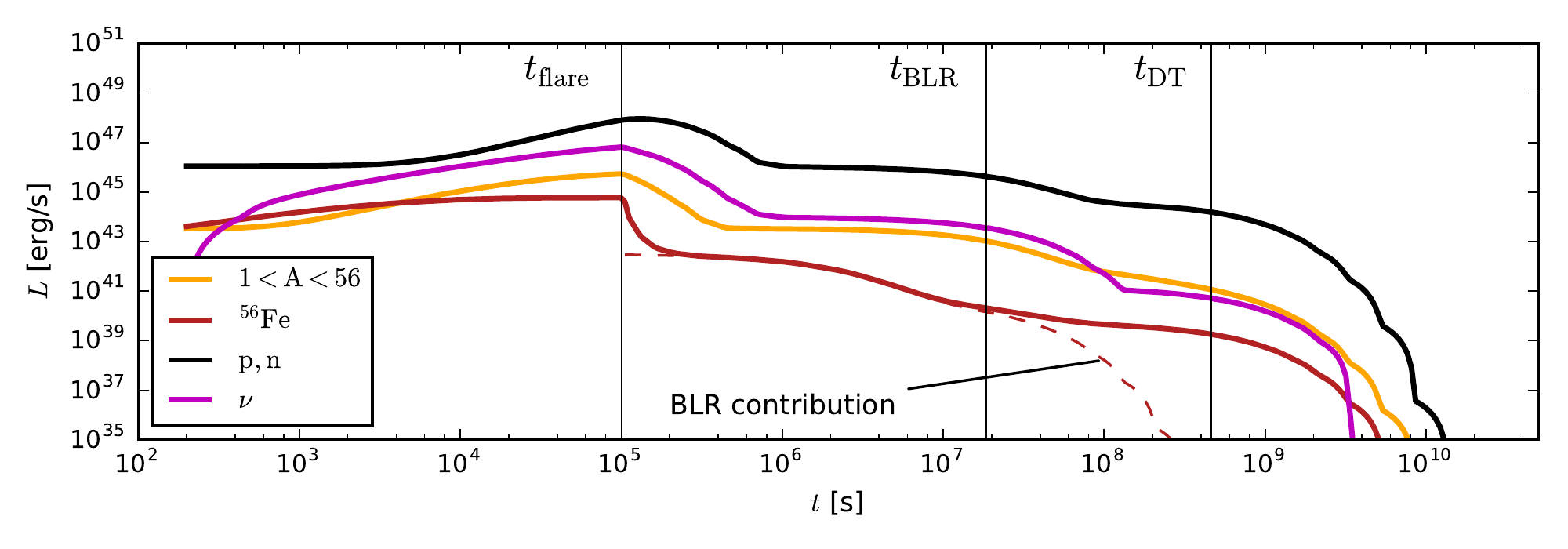} 
	\caption{Time evolution of the CR and neutrino luminosities escaping from each of the zones of the HL-FSRQ model, for iron-56 injection in a source with $L_\gamma=10^{48.8} \rm{erg/s}$ under the diffusive escape assumption. At $t_{\rm flare}$ the iron-56 injection halts and the decay of the spectra extends over a much longer timescale because of the larger volume.}
	\label{fig:lightcurves_diffusion}
\end{figure*}

We couple the BLR and DT zones in the numerical calculation by re-injecting the CRs that escape from the jet into the BLR, and then subsequently into the DT (see \figu{model} and \sect{methods})). The solver starts by integrating the blob system up to $\sim 50 \, t'_{\rm flare}$, while the injection of nuclei is switched off at $t'_{\rm flare}$ (and the electrons are assumed to maintain the radiation field). The CRs that escaped from the blob during the integration time are injected into a new set of equations describing the radiation environment of the BLR.
There we solve an initial value problem in the black-hole frame, starting with the CR densities $N_i(E, t = t_{\rm flare}) \neq 0$ ejected from the previous zone, using a homogeneous set of equations without primary nuclei injection. The DT region is treated similarly, starting with the BLR ejecta. The CR densities from the jet have to be boosted into the black-hole frame and rescaled to the new volume of the region.  Since no further injection takes place in the outer regions, the CR densities decrease over the light-crossing timescale of the respective zone through free-streaming escape and interactions with the external photon fields.

The instantaneous luminosity of the emitted CR densities (integrated over energy) is shown in \figu{lightcurves_diffusion} (for diffusive escape from the blob). Note that the time axis is logarithmic and the timescales for the outer zones are larger by several orders of magnitude. At the transition between jet and BLR zone one can clearly see the impact of the decay or outflow of in-source densities into the BLR. Although the interaction rates are smaller in the outer zones, they contribute sizably to the final result due to the much longer time scales.

The result does not change if CR emission is beamed at the blob-BLR transition, where the emission would be distributed over a smaller volume but experience the same target density. However, the beaming may be important at the BLR-DT transition; here the geometry of the DT has to be such that the jet direction traverses the radiation field of the torus (which is not limited to the torus, but drops gradually outside its boundaries). We will show later that the DT contribution does not qualitatively affect our results, although neutrino production in the DT might be relevant in some cases. 

\begin{figure*}[tp]
	\includegraphics[height=0.87\textheight]{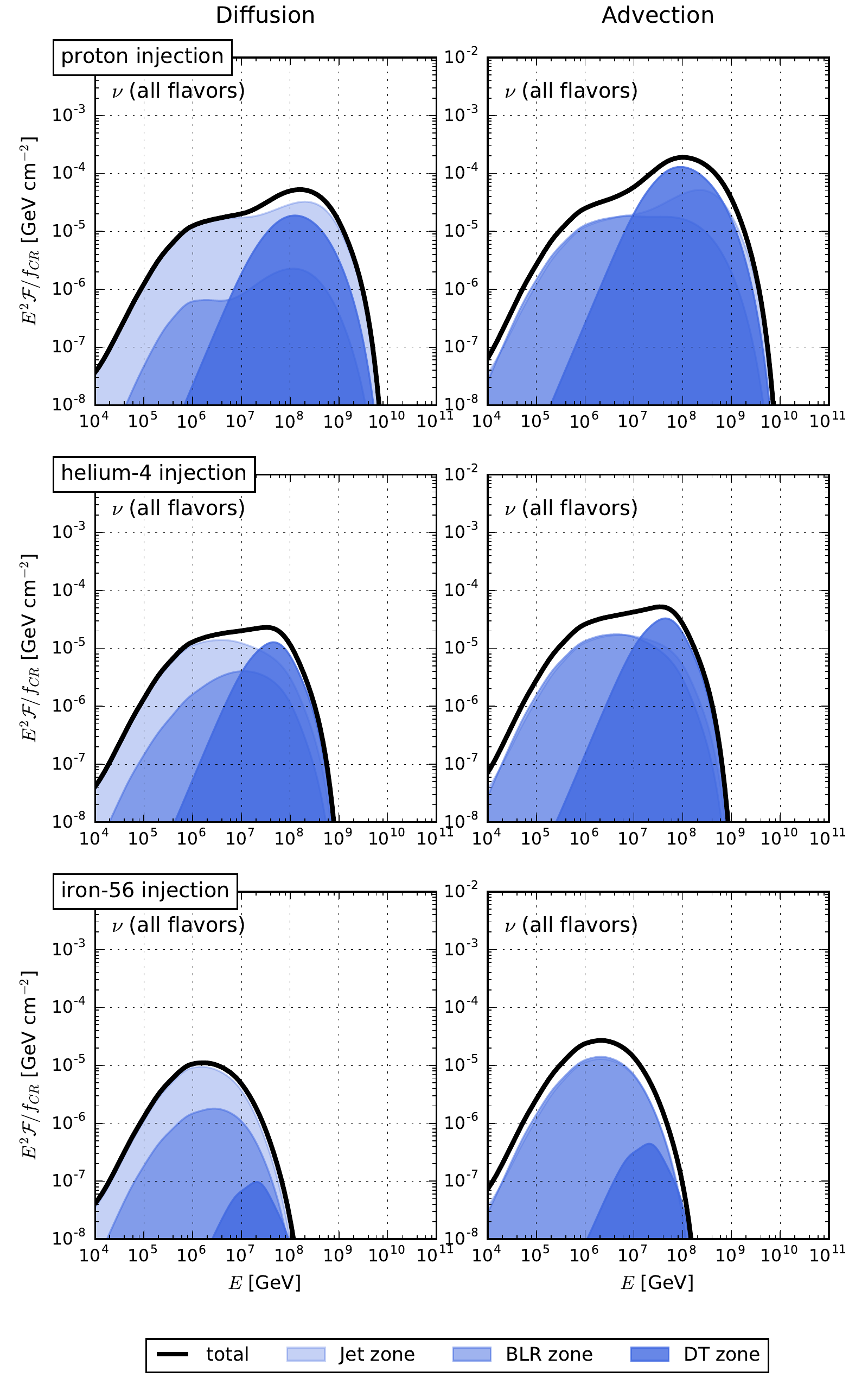}
	\caption{Ejected all-flavor neutrino fluence (in the observer's frame) for the injection of a pure composition of protons \textit{(top)}, helium-4 \textit{(middle)} and iron-56 \textit{(bottom)}, for an FSRQ with $L_\gamma=10^{48.8}$ erg/s (see \protect{\Tab~\ref{tab:sequence}}) at $z=1$, assuming the CR escape mechanism to be diffusive \textit{(left)} and advective \textit{(right)}. The contributions from the different regions are plotted separately. Note that the fluences from the three blazar zones add up to the total ejected spectrum, since neutrinos free-stream out of the source.
}
	\label{fig:zones_nu_fluence_49}
\end{figure*}

The neutrino spectra emitted by a HL-FSRQ are shown in \figu{zones_nu_fluence_49} for the two escape assumptions discussed earlier. The neutrino fluences from each zone are shown separately and have to be added to obtain the total. The corresponding CR spectra are discussed in \App~\ref{app:cr}. Since in the advection case more lower-energy CRs can escape the jet (having softer spectra), the BLR is populated with a higher initial density at $t = t'_{\rm flare}$. Therefore, a higher number of primaries is available for neutrino production in the outer zones compared to the diffusion case (darker blue shaded regions \figu{zones_nu_fluence_49}). For the pure proton injection (upper right panel) and advective escape, the outer zones (in particular the DT) notably broaden the peak at tens of PeV and increase the flux by factor $\sim 2$ compared to diffusive escape. This result is found to be consistent with previous studies~\cite{Murase:2014foa}.

The injection of helium-4 (middle panels) results in a similar behavior, except that the jet contribution is smaller as already discussed in \sect{neutrinos}. For the same reason as in the proton case, the advection case produces more neutrinos. Since the BLR is optically thick for heavier nuclei, most of the secondaries from iron-56 injection (lower panels) are absorbed in the BLR, such that neutrino emission from the DT is lower compared to the other two cases. Since HL-FSRQs clearly lie inside the nuclear cascade regime, they emit mostly nucleons and photo-disintegrated secondary nuclei. 

Independent of the injection composition, the escape through advection 
produces more neutrinos since more CR are injected in the BLR, which contributes 
at the same level as the production in the jet. Although the IR photon density is equal in the BLR and the DT (\figu{zones_sed}), the latter zone is more relevant for neutrino production for proton or helium-4 injection because of its much larger size -- which increases its optical depth to photo-pion production compared to the BLR (see \figu{zones_rates}). 

\section{Neutrinos and UHECRs evolving over the blazar sequence}
\label{sec:sequence}

\begin{table*}[t]
	\begin{tabular}{c|cc|cccc|cc}
		\hline
		~~ ID ~~ & Type & N       & $L_{\gamma}$      & $B'_{\mathrm{jet}}$ & $r_{\text{BLR}}$ & $r_{\text{DT}}$  & $L_{\nu}/L_{\gamma}$ & $L_{\nu}/L_{\gamma}$ \\
		~        & ~    & in 3LAC & $\log_{10}$,~erg/s & mG         & pc        & pc               & p                    & Fe \\
		\hline
		11 &  \multirow{3}{*}{HL-FSRQ}              & \multirow{3}{*}{49}      & $50.3$       & 5000 & 2.54  & 64    & $1.7\times10^{-1}$  & $8.7\times10^{-2}$  \\
		10 &  ~                                   & ~                        & $49.6$       & 2300 & 0.73 & 16    & $2.4\times10^{-1}$  & $4.7\times10^{-2}$  \\
		9 (B)  &  ~                                   & ~                        & $48.8$       & 900  & 0.18 & 4.0   & $1.5\times10^{-1}$  & $2.5\times10^{-2}$  \\
		\hline
		8  &  \multirow{2}{*}{~~LBL \& FSRQ~~} & \multirow{2}{*}{384}     & $47.7$       & 260  & --   & --    & $1.2\times10^{-2}$  & $1.2\times10^{-4}$  \\
		7  &  ~                                   & ~                        & $46.1$       & 40   & --   & --    & $4.2\times10^{-4}$  & $1.8\times10^{-5}$  \\
		\hline
		6  &  \multirow{3}{*}{~~IBL \& FSRQ~~} & \multirow{3}{*}{285}     & $45.7$       & 28   & --   & --    & $2.1\times10^{-4}$  & $2.1\times10^{-5}$  \\
		5  &  ~                                   & ~                        & $45.5$       & 21   & --   & --    & $8.6\times10^{-5}$  & $1.8\times10^{-5}$  \\
		4 (A)  &  ~                                & ~                        & $44.6$       & 7.1  & --   & --    & $3.3\times10^{-6}$  & $1.2\times10^{-6}$  \\
		\hline
		3  &  \multirow{3}{*}{HBL}        & \multirow{3}{*}{29}      & $43.5$       & 2.2  & --   & --    & $1.1\times10^{-7}$  & $4.1\times10^{-8}$  \\
		2  &  ~                                   & ~                        & $42.5$       & 0.69 & --   & --    & $3.9\times10^{-9}$  & $1.4\times10^{-9}$  \\
		1  &  ~                                   & ~                        & $41.6$       & 0.23 & --   & --    & $1.4\times10^{-10}$ & $4.7\times10^{-11}$ \\
		\hline
	\end{tabular}
	\caption{Models and parameters considered in this work. The third column shows the number of observed blazars with measured redshift in Fermi-3LAC catalog \cite{Ackermann:2015yfk}. Note that among model ID 4 to 8, there are large overlaps between the FSRQ and BL Lac populations. From model ID 1-8, we ignored the BLR and DT zones, since they are either negligible in all BL Lacs or the blob in the jet lies outside the regions in those FSRQs. The last columns show the results from the simulations (obtained neutrino-to-gamma-ray luminosity ratio for a CR loading $f_{\rm CR} = 10$ and considering diffusive CR escape).}
	\label{tab:sequence}
\end{table*}

It has long been proposed that the peak frequencies of the blazar SEDs are anti-correlated with their bolometric luminosities, which is referred to as ``blazar sequence''~\cite{Fossati:1998zn}. This trend becomes more prominent from the recent analysis \cite{Ghisellini:2017ico} of a sample of 747 blazars with measured redshifts from the Fermi 3LAC catalog, compared to the small sample of merely 33 EGRET blazars in~\cite{Fossati:1998zn}. Despite the debate on whether this is influenced by a selection bias due to the limitation of the sensitivity of the observation instruments, this phenomenological prescription does adequately describe at least the sample of the {\it observed} blazars. Therefore it provides us with simplified and representative SEDs in each luminosity bin. We list the used luminosity bins in \Tab~\ref{tab:sequence}.

\begin{figure}[tb!]
	\includegraphics[width=\columnwidth]{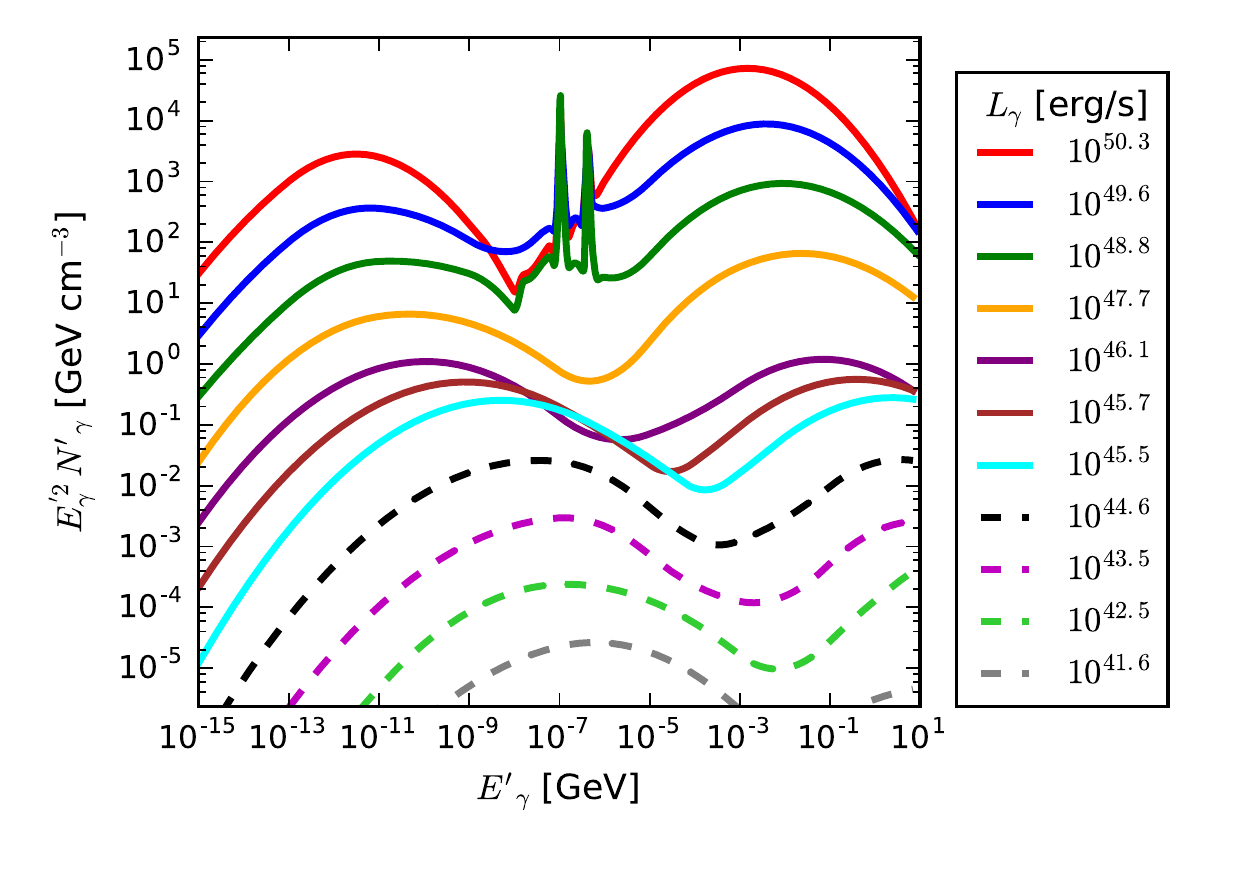}
	\caption{Spectral Energy Distributions (SEDs) used in this study, following the original blazar sequence~\protect{\cite{Fossati:1998zn,Ghisellini:2017ico}}. The solid lines are taken from~\protect{\cite{Murase:2014foa}}, while the dashed lines are extrapolated SEDs representing the extension of the sequence to high-synchrotron-peaked BL Lacs. Note that the SEDs relevant for the secondary production are shown, whereas the observed SEDs for some FSRQs can contain additional external contributions visible in photon observations (see main text for details).}
	\label{fig:sequence_sed}
\end{figure}

The higher luminosity bins are mostly populated by FSRQs from observations, as in models 7-11 of \Tab~\ref{tab:sequence}. In our scenario, an emitting blob of a fixed radius and distance to the central black hole is assumed, but all the physical quantities in the table increase with luminosity. Therefore, the blob can be located inside (see Table \ref{tab:sequence}, model 7-8) or outside the BLR and dusty torus (Table \ref{tab:sequence}, model 9-11; see discussion in \sect{geometry}). The corresponding target photon SEDs  in the blob frame are shown in the top five curves of \figu{sequence_sed} for models 7-11, respectively. In particular, for the top three curves (models 9-11), the contributions from the BLR and DT are clearly seen, on top of the smooth two-hump structure (photons produced from the blob internally). For models 7 and~8, the contributions of the external photon fields are (by construction) not present in the interactions. The lower luminosity bins are mainly populated by BL Lacs, corresponding to models 1-6 of Table \ref{tab:sequence} and the bottom six curves of \figu{sequence_sed}, where the BLR and dusty torus contributions are absent. To account for the sub-threshold low-luminosity blazars, we roughly extrapolated the blazar sequence down to $L_{\gamma}\sim10^{42}$ erg/s (models 1-4). 

We now apply our model to the sources of the blazar sequence. For low-luminosity sources only the radiative processes in the jet will be considered, given the geometry arguments presented before, while the simulation of HL-FSRQs will follow the three-zone model introduced in~\sect{fsrq}.

\begin{figure}[tb!]
	\includegraphics[width=\columnwidth]{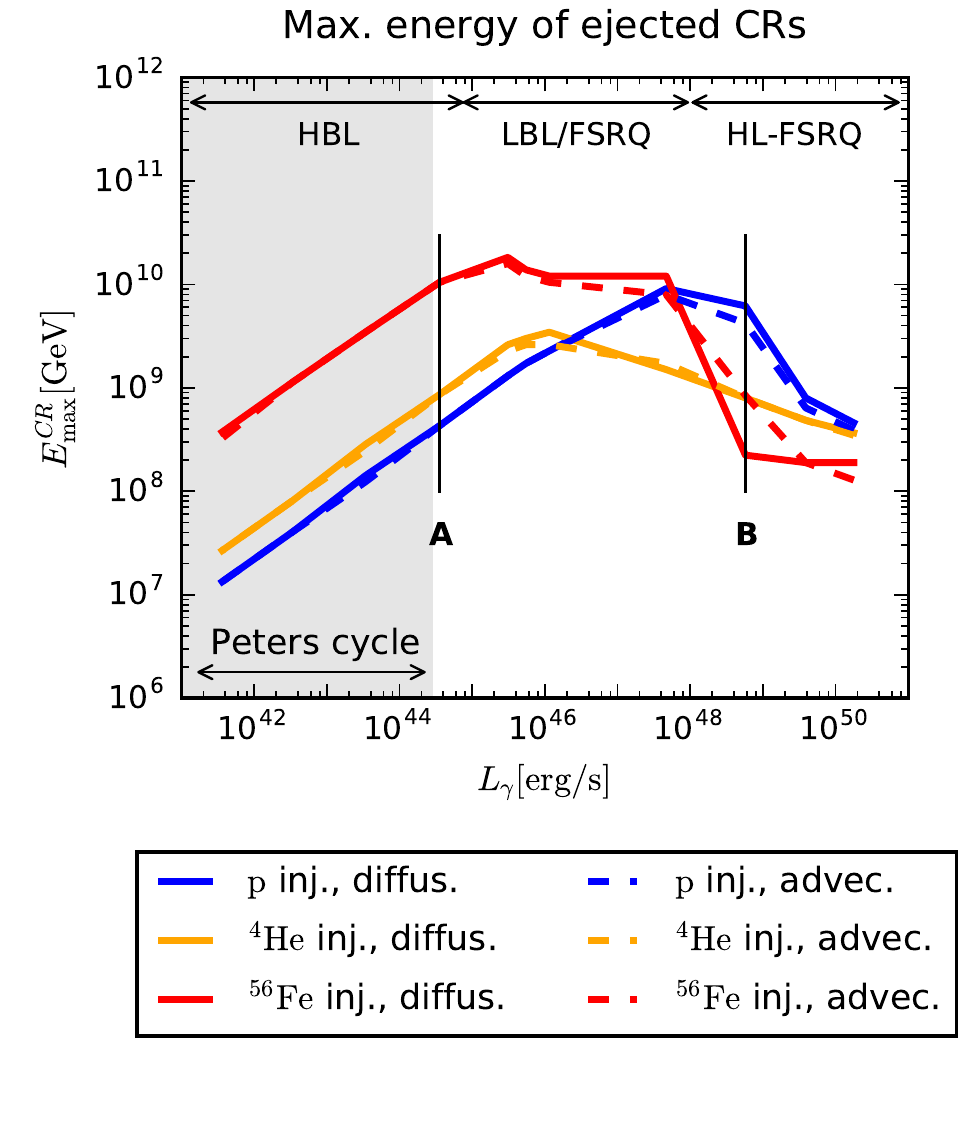}{\vspace{-6mm}}
	\caption{Maximum ejected CR energy for each source described in \protect{\tabl{sequence}} and \protect{\figu{sequence_sed}}. Only BL Lacs emit CRs that follow the Peters cycle $E_{\rm max} \sim Z$, \ie~a scaling of the maximum energy with $Z$. The luminosity marked \textit{A} corresponds to the nuclear survival prototype from \Sec~\ref{sec:nuclear_regimes}, while \textit{B} corresponds to the HL-FSRQ discussed in \Figs~\ref{fig:zones_sed} to~\ref{fig:zones_nu_fluence_49}.}
	\label{fig:peters_cycle}
\end{figure}

Figure~\ref{fig:peters_cycle} shows the maximum CR energy ejected by each source, for different injected isotopes and for either of the escape assumptions discussed in \sect{escape}. For HBLs, where photo-nuclear/-hadronic interactions are sub-dominant up to the highest injection energies, the maximum energy is determined from the Hillas condition in the blob. As a result, we have $E_\mathrm{max} \propto B^{\prime} \propto L_{\gamma}^{1/2}$ along the blazar sequence, since a constant energy partition $\epsilon_\mathrm{B}$ is assumed for the magnetic fields. In the range indicated by the gray band, the maximum energy is proportional to the charge of the injected isotope, which corresponds to the requirement that cosmic accelerators follow the Peters cycle \cite{peters1961} (rigidity-dependent maximum energy), an often implied relationship in textbook physics such as the Hillas Plot and in recent phenomenological studies, \eg\ \cite{Aab:2016zth}. This relationship breaks down at the transition between the nuclear survival and nuclear cascade regimes at $L_\gamma \gtrsim 10^{45}-10^{46}$ erg/s, where photo-hadronic interactions become efficient and limit the maximum energy of ejected CRs. A further increase of the luminosity results in a reduction of the maximum energy for the reasons discussed in \sect{nuclear_regimes}. The effect is more pronounced for heavier injection isotopes, as we can see from the three cases displayed in \figu{peters_cycle}. Note that the maximum energy is not significantly affected
by the escape scenario.

\begin{figure*}[tb!]
	\includegraphics[width=0.9\textwidth]{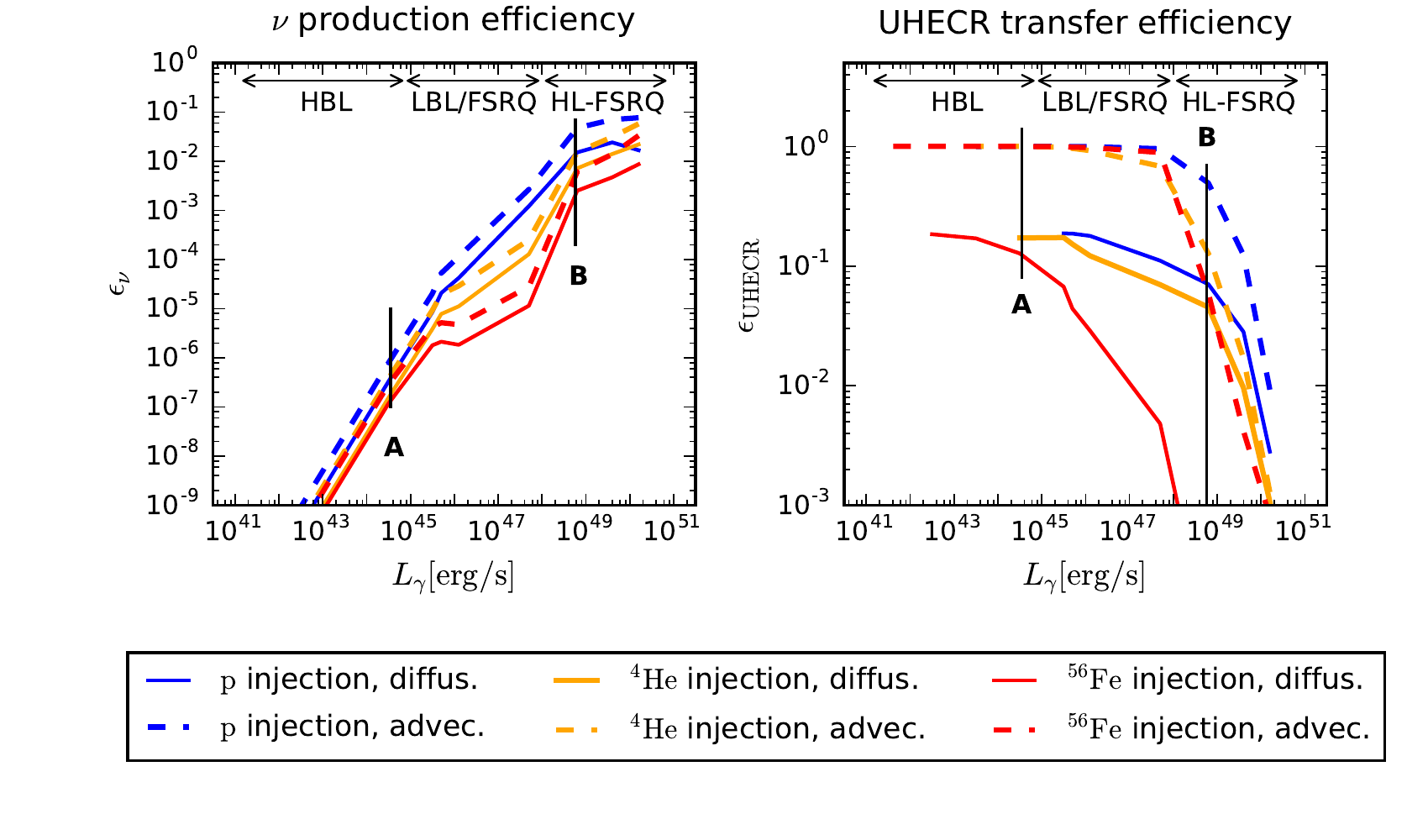}{\vspace{-3mm}}
	\caption{\textit{Left:}  All-flavor neutrino production efficiency for blazars of different luminosities, following the blazar sequence described in \protect{\tabl{sequence}} and \protect{\figu{sequence_sed}}. \textit{Right:} Cosmic-ray energy transfer efficiency in the UHE range ($E_{\text{CR}}>10^{9}$ GeV in the black-hole frame) for the same sources. The colors represent three different pure injection compositions; solid and dashed curves represent diffusive and advective escape, respectively. The luminosity marked \textit{A} corresponds to the nuclear survival prototype from \Sec~\ref{sec:nuclear_regimes}, while \textit{B} corresponds to the HL-FSRQ discussed in \Figs~\ref{fig:zones_sed} to~\ref{fig:zones_nu_fluence_49}.}
	\label{fig:transfer_efficiency}
\end{figure*}

We now turn to the production efficiency of UHECRs and neutrinos across the blazar sequence. We define the efficiency in the production of UHECRs as the ratio of the total energy of the time-integrated spectrum of all ejected CR species above $10^9 \rm{GeV}$ in the black-hole frame, and the total injected energy in CRs above the same energy as 
\begin{equation}
	\epsilon_\text{CR} \equals
        \frac{\sum\limits_\text{all CRs}\int_0^\infty \int\limits_{10^9 \mathrm{GeV}} Q^\prime_{\text{esc}} (E^\prime,t^\prime) \, dE^\prime \, dt^\prime}
             {\int_0^\infty \int\limits_{10^9 \mathrm{GeV}}{Q^\prime_\text{inj}}(E^\prime,t^\prime) \,  dE^\prime \, dt^\prime} \, ,
	\label{equ:efficiency}
\end{equation}
where $Q^\prime_{\text{inj}}$ is the injection rate from the acceleration zone of the injected isotope, as defined in \equ{inj}. Similarly, the neutrino production efficiency $\epsilon_\nu$ is obtained by performing the sum in the numerator over all neutrino species, and both integrals over all energies. 

This quantity does not depend on the CR loading factor $f_{\text{CR}}$ of the sources, which cancels out in \equ{efficiency} (as opposed to the numbers listed in the last columns of \Tab~\ref{tab:sequence}, which are computed for $f_{\text{CR}}=10$).

The neutrino production efficiencies are shown in the left panel of \figu{transfer_efficiency}. Clearly, HL-FSRQs can be identified to be the best neutrino emitters of the blazar sequence, with $\epsilon_\nu\sim10^{-3}$ to $10^{-1}$. For less luminous blazars the photon field densities are too low for efficient neutrino production, with HBL Lacs being the least efficient neutrino sources. From the same plot we can see that HBL Lacs follow the expected relation $L_\nu\sim\epsilon_\nu L_\gamma\sim L_\gamma^2$; above $L_\gamma=10^{45}$ erg/s, this scaling becomes weaker, and beyond $L_\gamma=10^{49} \mathrm{erg/s}$ it becomes $L_\nu\sim L_\gamma$. There is hardly any dependence on the UHECR escape mechanism, apart from the $\sim2$ times more efficient production for the advective case, as discussed in \sect{fsrq}. 

On the right-hand side panel of \figu{transfer_efficiency} we plot the UHECR transfer efficiency. The overall tendency is the opposite of the neutrino efficiency: HL-FSRQs cannot efficiently emit UHECRs due to strong photo-hadronic interactions and also to the additional radiation zones that suppress the UHECR part of the spectrum, while lower-luminosity objects such as LL-FSRQs and BL Lacs may allow for more efficient CR survival and escape.  HBL Lacs are expected to be efficient sources of UHECR regardless of the injected isotope, since 100\% of the injected UHECRs are emitted. In the advective escape scenario (dashed curves in \figu{transfer_efficiency}) blazars become efficient UHECR source candidates. Note, however, that the spectra of emitted CR are generally softer. For diffusion-dominated CR escape, the production efficiency is low. The UHECR transfer efficiency should be interpreted together with the maximum achievable energy in (see \figu{peters_cycle}): While, for example, at low energies a large fraction of the injected energy is transferred, the magnetic field is too low to allow for very high UHECR energies of light isotopes. Therefore, each injection isotope  has a ``sweet spot'' where UHECR can be accelerated to sufficiently high energies and escape from the source environment. For example, for helium injection,  this sweet spot seems to be around $10^{46} \, \mathrm{erg \, s^{-1}}$. If, in addition, neutrinos are to be efficiently produced,  higher luminosities are preferred; $L \sim 10^{48} \, \mathrm{erg \, s^{-1}}$ seems to be a good trade-off between neutrino and UHECR production. 

It is worth noting that in \cite{Ghisellini:2017ico}, with the inclusion of 747 blazars from the Fermi-3LAC catalog, the blazar sequence becomes more prominent and the SEDs for each luminosity bin can be constructed with more refined features. In comparison with the blazar sequence considered in this work, the main differences appear in the gamma-ray slopes and in the luminosity ratio of the second to the first hump (Compton dominance), in particular for low-luminosity blazars. Since the overall shapes of the SEDs are similar, as demonstrated in Fig. 9 of \cite{Ghisellini:2017ico}, we do not expect significant changes of the neutrino and CR spectra obtained in this work.

\section{Summary and conclusions}
\label{sec:conclusion}

We have studied blazars as sources of UHECRs and neutrinos including the injection of isotopes heavier than hydrogen into the jet. We have identified two important regimes, depending on jet luminosity and size of the blob: In the nuclear survival regime, corresponding to low luminosities or large blob sizes, the source is found to be optically thin to photo-nuclear interactions. Neutrinos are mostly produced due to photo-meson production off the primary nuclei, and the maximum CR energies behave as expected from Peters cycle, \ie~$E_{\rm max} \sim Z$. In the nuclear cascade regime, corresponding to high luminosities or small blob sizes, the source is optically thick to disintegration of the injected nuclei at the highest energies, and the nuclear cascade becomes  more populated. Neutrinos are copiously produced in photo-meson processes off the secondary nuclei generated in the nuclear cascade, and the maximum CR energies saturate due to the photo-nuclear processes. Compared to GRBs, the nuclear cascade ceases faster with decreasing mass number because the disintegration rate depends more strongly on energy. In addition, the neutrino peak flux and its energy depend on the injected nuclear composition, where heavy injection compositions are more in favor of observed data. 

The evolution of neutrino and CR production efficiencies has been studied over the blazar sequence including the corresponding SEDs of target photon spectra based on observations. In order to have a more refined model for HL-FSRQs, for which the jet production region may lie within the BLR, we have included the external radiation fields in the jet and the propagation of CRs through the radiation fields of the BLR and DT radiation fields. Our main conclusion is that the neutrino production efficiency strongly increases with luminosity from an extremely low value for HBLs, over LBLs/FSRQs, and then saturates for HL-FSRQs. For HBLs, we observe a scaling $L_\nu \propto L_\gamma^2$, whereas for HL-FSRQs, we find $L_\nu \propto L_\gamma$. On the contrary, the UHECR transfer efficiency is high for HBLs, then decreases for LBLs/FSRQs, and it is low for HL-FSRQs. This opposite behavior is expected since the strong radiation fields lead to photo-nuclear disintegration and effective neutrino production at the same time. If in addition the maximum achievable CR energy is taken into account, it seems that the optimal luminosity for UHECR acceleration and escape is around $L_\gamma \sim 10^{46} \, \mathrm{erg \, s^{-1}}$ for isotopes heavier than helium. 
The precise level of disintegration can be somewhat controlled with the blob size, which is typically a parameter coming from self-consistent radiation models. 

Note that while our observations hold for individual sources, the contribution of HBLs vs. LBLs vs. FSRQs to the diffuse CR and neutrino fluxes depends on the luminosity and redshift distribution functions of the sources. However, for individual sources there may be a ``sweet spot'' in the LBL/FSRQ range around $L_\gamma \sim 10^{48} \, \mathrm{erg \, s^{-1}}$ (for helium injection), where both UHECRs and neutrinos are efficiently produced -- which corresponds to high-luminosity BL Lacs and intermediate-luminosity FSRQs. In order to test the origin of CRs with neutrinos, this may be the region to search for correlations. Furthermore, note that we use a one-zone model for the jet, whereas multi-zone models may lead to an enhanced neutrino production for BL Lacs by enhancing the radiation fields from the relative motion of the zones~\cite{Tavecchio:2014iza,Tavecchio:2014eia}. However, in these cases the level of photo-nuclear disintegration will be enhanced as well, which means that the anti-correlation between neutrino and CR production continues to hold.

On a side note, we have also discussed the impact of the UHECR escape mechanism from the jet, where we have focused on a diffusion scenario, leading to hard ejection spectra motivated by recent Auger fits, and an advection scenario, leading to unmodified  escape at lower energies (compared to the acceleration spectrum) and motivated by the physics of collimated jets without adiabatic cooling. While we have not found qualitative differences in our observations for neutrinos versus CRs over the blazar sequence, the ejected CR spectrum and composition, as well as the transfer efficiency to UHECRs can be quantitatively very different. Both spectrum and composition of the diffusion scenario agree better with recent Auger observations, whereas the advection scenario leads to more efficient UHECR ejection. The overall neutrino production is hardly affected by the escape assumption, but we observe a significant contribution of neutrino production in the BLR of HL-FSRQs for advective escape, which is not present for diffusive escape. Note that since the assumed diffusive escape mechanisms is rather conservative, and the advective escape assumption quite aggressive, we expect that any realistic scenario should in fact lie in between.

We conclude that HL-FSRQs are very efficient neutrino emitters, whereas UHECRs from blazars will efficiently come from low- and intermediate-luminosity objects. These conclusions hold in the one-zone model for the jet zone with target photons derived from observed SEDs. 
These observations may be especially relevant for individual source correlation studies and catalog stacking searches, whereas the contributions to the diffuse flux require the input from the luminosity distribution. For example, in spite of their low neutrino luminosity, a sufficiently large number of LBLs which are not detectable otherwise can still dominate the diffuse neutrino flux. Another variable is the CR loading, which may vary for different blazar families or evolve over the blazar sequence. Further studies will reveal if the UHECR and neutrino fluxes can be powered at the same time, and how the recent stacking results for blazars need to be interpreted.

{\bf Acknowledgments.}  We would like to thank Kohta Murase for fruitful discussions.

This project has received funding from the European Research Council (ERC) under the European Union’s Horizon 2020 research and innovation programme (Grant No. 646623).

\bibliographystyle{apsrev4-1}

\clearpage
\newpage

\begin{appendices}

\section{Ejected CR spectra in HL-FSRQs}
\label{app:cr}
\begin{figure*}[tpb!]
	\includegraphics[height=0.8\textheight]{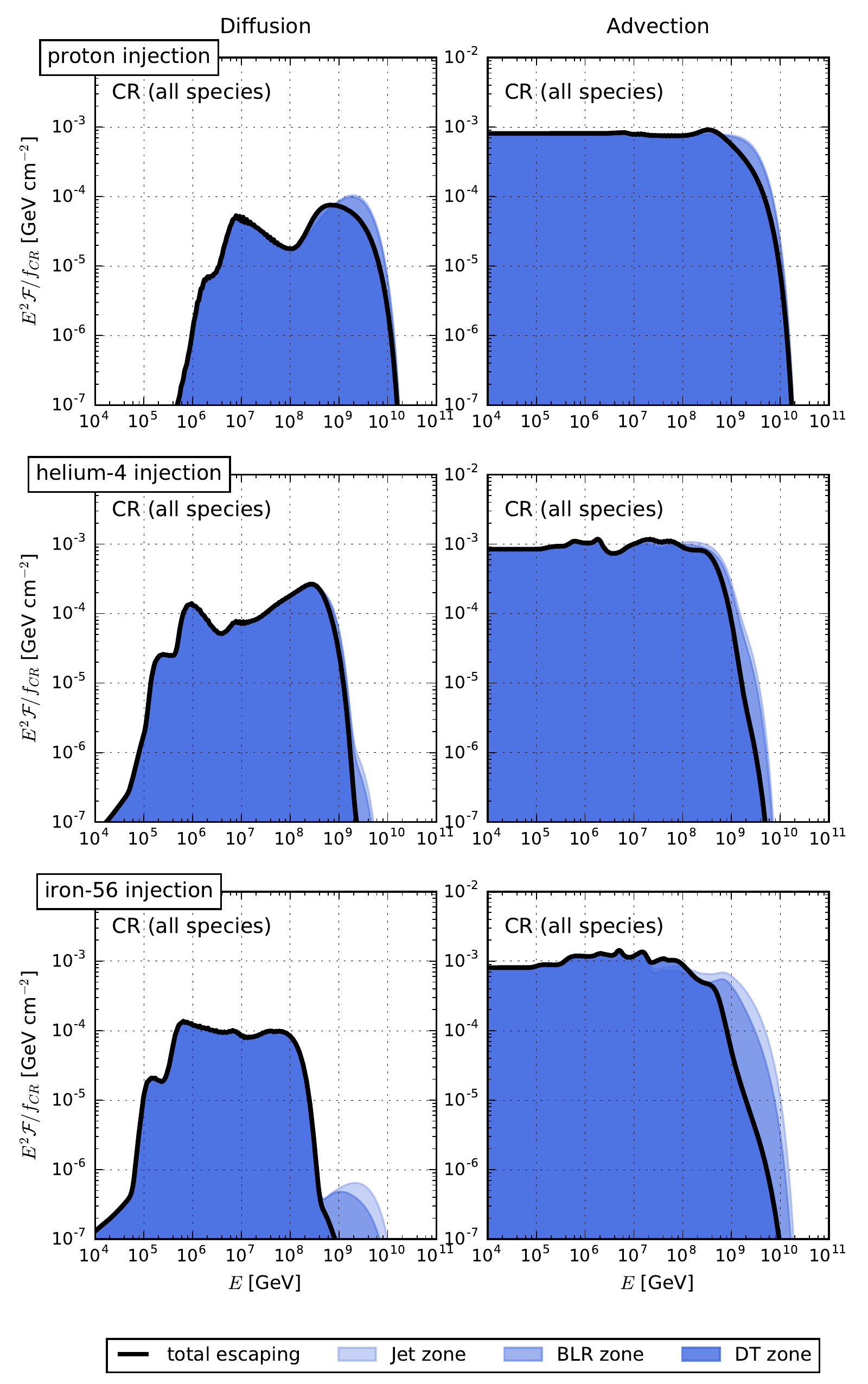}
	\caption{Ejected CR fluence for the injection of a pure composition of protons \textit{(top)}, helium-4 \textit{(middle)} and iron-56 \textit{(bottom)}, for a HL-FSRQ with $L_\gamma=10^{48.8}$ erg/s (see \protect{\Tab~\ref{tab:sequence}}), assuming the CR escape mechanism to be diffusive \textit{(left)} and advective \textit{(right)}. The effects of the different regions are plotted separately, \ie~the escaping spectrum from the respective region is shown, and the final escape spectrum is shown as the black curve. Note that the additional zones reprocess the CRs and suppress the ejected spectrum, and that only those particles ejected from the last zone will enter the circum-AGN medium. A redshift $z=1$ was assumed for the fluence calculation, which is shown in the observer's frame. The energies are expressed in the observer's frame as well, although the interactions during the CR propagation are not taken into account.}
	\label{fig:zones_cr_fluence_49}
\end{figure*}

In this section we discuss the effect of the external zones of a HL-FSRQ and the escape assumption on the ejected CR spectrum. In \figu{zones_cr_fluence_49} the boarders of the shaded regions represent the total cosmic ray fluence transferred between the successive zones. Contrary to the behavior in neutrinos (compare with \figu{zones_nu_fluence_49}), the CR fluence is reduced due to photo-nuclear losses in the outer zones. As already discussed in \figu{transfer_efficiency}, the diffusive escape assumption results in an approximately ten-fold lower CR output from the jet zone, which is even further suppressed at high energies during propagation through the BLR and the DT (compare left- and right-hand panels). The net impact of the outer zones is therefore smaller for diffusion, in agreement with the behavior of the ejected neutrino spectra. In the advection case this impact is more pronounced for heavier injection composition, since the optical thickness to photo-nuclear interactions is higher for nuclei. The outer zones do not affect the general conclusion about the spectral index, which remains softer for advection compared to diffusion.

\section{Other assumptions on magnetic field and acceleration efficiency}
\label{app:acceleration_efficiency}

We have so far assumed that the energy budget of the magnetic field corresponds to a constant fraction of the total energy of the jet. It is the simplest assumption since we do not explicitly compute the components of the SED, as explained in the last paragraph of \sect{jet_model}. In this appendix we explore parameter ranges that are closer to those considered in SSC models that are able to generate the SEDs (\eg\ \cite{Tavecchio:2009zb,Dermer:2013cfa}, where the hadronic components are sub-dominant). In such models, the ratio of the energy densities $u_\mathrm{B}$ to $u_\mathrm{syn}$ can be immediately derived from the SSC parameter $Y_\mathrm{SSC}$, which roughly corresponds to the luminosity ratio of the second hump to the first hump in the observed SED. Then, the magnetic field strength is seen to lie in the range $0.1-1 \,\mathrm{G}$ for BL Lacs and $B\sim \text{a few}\, \mathrm{G}$ for FSRQs. In addition, studies of diffusive shock acceleration \cite{Inoue:2016fwn} suggest a lower value for the acceleration efficiency compared to our choice of $\eta = 1$ (see \equ{acceleration_efficiency}).

The values of $B$ derived from SSC models can be well described by a power-law relation $B'\sim L_\gamma^{1/5}$ along the blazar sequence (in contrast to $L_\gamma^{1/2}$ as in the main text, which yields much lower magnetic fields for BL Lacs). This results in a running value of the energy partition factor $\epsilon_B=\epsilon_B(L_\gamma)$ instead of a constant one. As a consequence, the dimmest HBL has a magnetic field strength of $B'=0.09\, \mathrm{G}$, while the brightest FSRQ has $B'=5 \, \mathrm{G}$, both consistent with the results from \cite{Tavecchio:2009zb,Dermer:2013cfa}. Furthermore, we now decrease the acceleration efficiency to $\eta=0.1$.

\begin{figure*}[htpb!]
	\includegraphics[width=\textwidth]{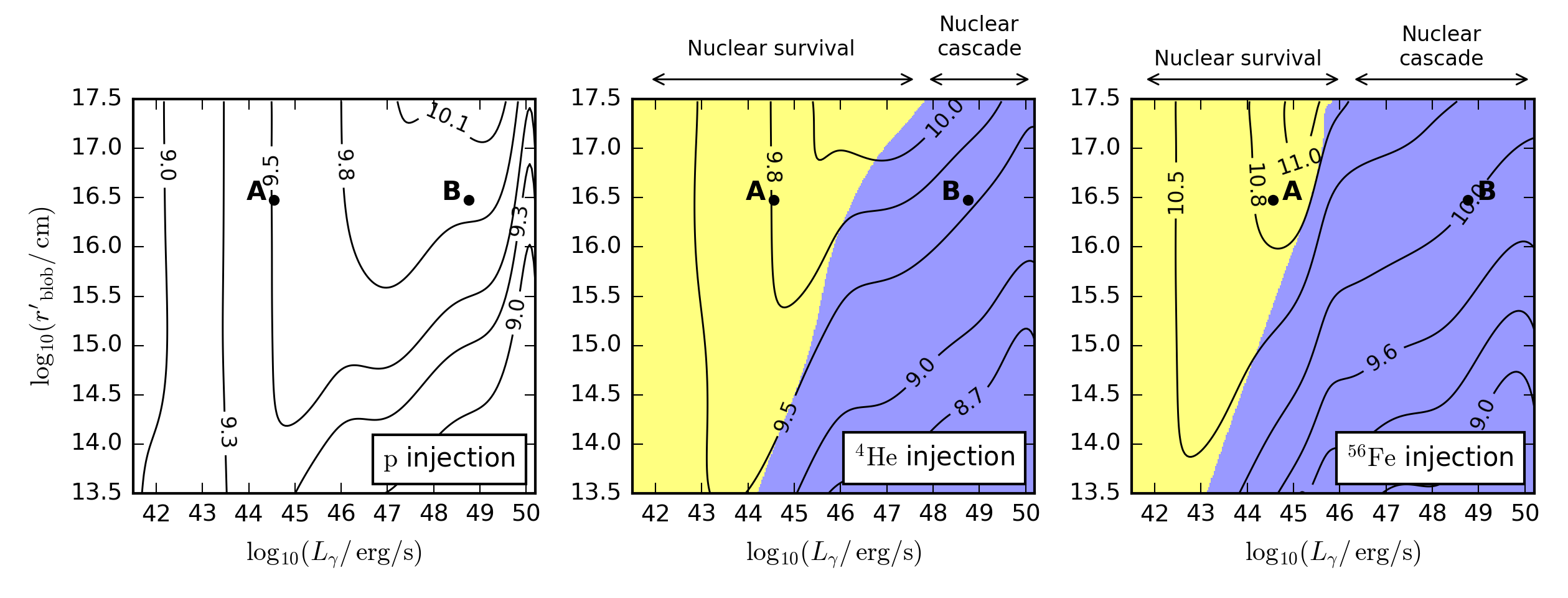}
	\caption{The contours indicate the maximum energy of the injected isotope, $\log_{10}{(E_\text{max}/\text{GeV})}$, and the colors represent the two nuclear regimes, as in \figu{parameter_scan}. Here, however, a lower acceleration efficiency is considered ($\eta = 0.1$ instead of $\eta = 1$), as well as a different magnetic field range, as described in the text of \App{}~\ref{app:acceleration_efficiency}.}
	\label{fig:app_parameter_scan}
\end{figure*}

\begin{figure*}[htbp!]
	\includegraphics[width=\textwidth]{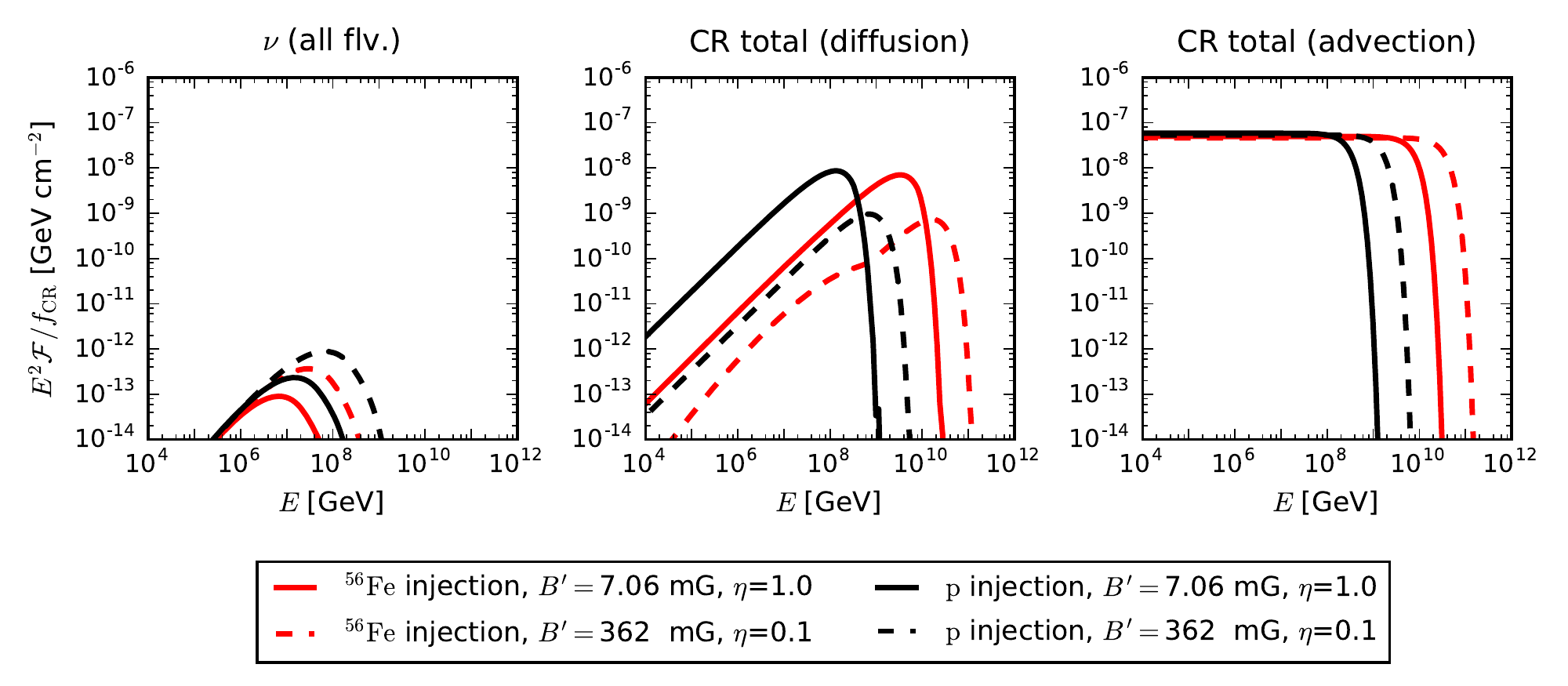}
	\caption{Ejected spectra of neutrinos \textit{(left)}, ejected CRs through diffusive escape \textit{(center)} and ejected CRs through advective escape \textit{(right)}, from a source in the nuclear survival regime ($L_\gamma=10^{44.6} \mathrm{erg/s}$). The solid lines are the spectra obtained with the default $\eta$ and magnetic field parameters from the main text, while the dashed lines are the spectra obtained with an acceleration efficiency of $\eta=0.1$ and magnetic field strength of $B'=0.362 \, \mathrm{G}$. The latter configuration favors CR production up to higher energies in BL Lacs due to the strong magnetic fields. A redshift $z=1$ is assumed for the fluence calculation, which is shown in the observer's frame.}
	\label{fig:app_b_eta}
\end{figure*}

In \figu{app_parameter_scan} we show the maximum injection energy of three isotopes under these conditions. For low-luminosity BL Lacs, in spite of the lower acceleration efficiency, the high magnetic fields lead to a higher maximum injection energy, regardless of the injected element (\cf~\figu{parameter_scan}). The separation between the nuclear survival and nuclear cascade regimes does not change noticeably. The increase in maximum energy is also observed in the ejected neutrino and CR spectra for low-luminosity blazars, as shown in \figu{app_b_eta}. The increased maximum CR energies in the jet lead to a ten-fold increase in ejected neutrinos. Recall, however, that blazars in the nuclear survival case are generally not very efficient neutrino sources (see \figu{transfer_efficiency}). On the other hand, for the ejected CRs the different parameter set leads to a flux decrease by one order of magnitude in the case of diffusive escape (see middle panel of \figu{app_b_eta}). That is because the Larmor radius cannot reach the size of the region, even at the highest energies, and therefore CRs are always affected by confinement. For advective escape, the normalization of ejected CR spectrum is not affected, since this escape mechanism does not depend on the magnetic field strength.

\begin{figure}[htbp!]
	\includegraphics[width=\columnwidth]{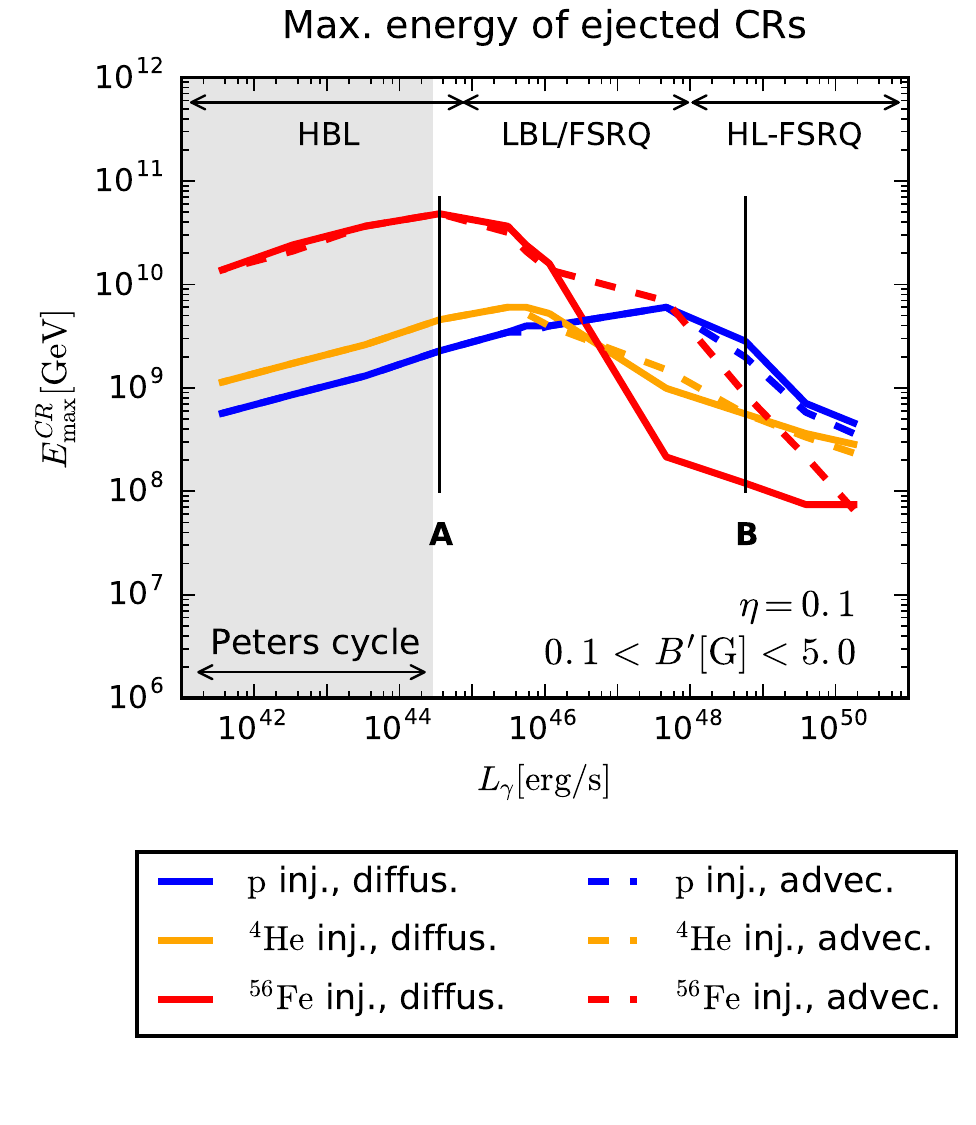}{\vspace{-6mm}}
	\caption{Maximum ejected CR energy for each source, as in \protect{\figu{peters_cycle}}, but assuming an acceleration efficiency ten times lower ($\eta = 0.1$) and magnetic fields in the sub-Gauss to Gauss range. With this parameter set the maximum energy of CRs emitted by HBL Lacs is more than an order of magnitude higher, regardless of the injected element.}
	\label{fig:app_peters_cycle}
\end{figure}

We have also studied the effect of this parameter set on the maximum ejected energy across the entire blazar sequence, as shown in \figu{app_peters_cycle}. We note that the qualitative conclusions drawn in \sect{sequence} are unchanged: there is an intermediate luminosity range where the ejected CR energy is maximal. The maximum ejection energy in the low-luminosity range is affected by the softer scaling of $E_\mathrm{max} \propto B^{\prime} \propto L_{\gamma}^{1/5}$. As a result, CRs ejected by HBLs can reach energies that are ten times higher compared to the result in \sect{sequence} (\cf~\figu{peters_cycle}). For iron-56 injection, the luminosity range in which the ejected CR reach the highest energies moves from LBL/FSRQ to HBL/LBL. This aspect, however, is not observed for helium-4 and protons. 

\begin{figure*}[htbp!]
	\includegraphics[width=0.9\textwidth]{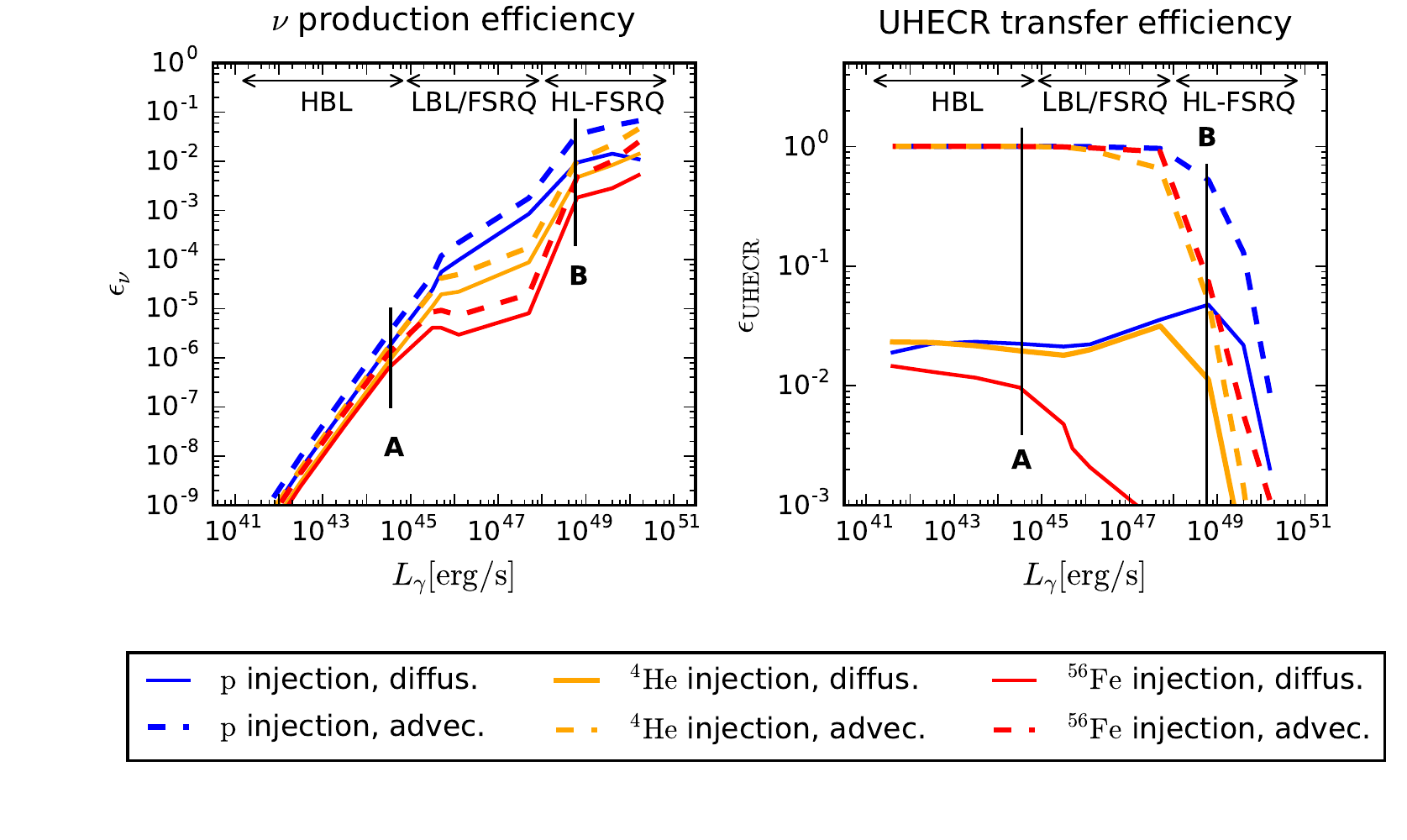}{\vspace{-6mm}}
	\caption{\textit{Left:}  All-flavor neutrino production efficiency for blazars of different luminosities, following the blazar sequence described in \protect{\tabl{sequence}} and \protect{\figu{transfer_efficiency}}, for the alternative values of magnetic field strength and acceleration efficiency discussed in \App{}~\ref{app:acceleration_efficiency}. \textit{Right:} Cosmic-ray energy transfer efficiency in the UHE range ($E_{\text{CR}}>10^{9}$ GeV in the black-hole frame) for the same sources and the same values of magnetic field and acceleration efficiency. The colors represent three different pure injection compositions; solid and dashed curves represent diffusive and advective escape, respectively. The luminosity marked \textit{A} corresponds to the nuclear survival prototype from \Sec~\ref{sec:nuclear_regimes}, while \textit{B} corresponds to the HL-FSRQ discussed in \Figs~\ref{fig:zones_sed} to~\ref{fig:zones_nu_fluence_49}.}
	\label{fig:app_transfer_efficiency}
\end{figure*}

Finally, in \figu{app_transfer_efficiency} we show the impact on the transfer efficiency (\cf{}\ \figu{transfer_efficiency}). While the conclusions are qualitatively unaffected, the modified parameter set results in a higher neutrino production efficiency of HBLs and an order-of-magnitude lower UHECR transfer efficiency of BL Lacs and low-luminosity FSRQs. On the other hand, under these conditions all BL Lacs down to the lowest luminosities are capable of ejecting CRs above $10^9\, \mathrm{GeV}$, as opposed to the parameter set discussed previously, where the low magnetic fields prevented HBLs from ejecting UHECRs in the case of a diffusive escape mechanism.

\end{appendices}

\end{document}